\documentclass[prd,twocolumn,showpacs,nofootinbib,floatfix,superscriptaddress]{revtex4}

\usepackage{amsfonts}
\usepackage{amsmath}
\usepackage{amssymb}
\usepackage{upgreek}
\usepackage{bm}
\usepackage{dcolumn}
\usepackage{epsfig}
\usepackage{graphicx}
\usepackage{graphics}
\usepackage{subfigure}
\usepackage{wrapfig}
\usepackage[latin1]{inputenc}
\usepackage{latexsym}
\usepackage{rotating}
\usepackage{float}
\usepackage{hyperref}
\usepackage{xspace} 
\usepackage[usenames,dvipsnames]{color}

\usepackage{ulem}
\definecolor {darkgreen}{rgb}{0.2,0.7,0.2} 

\newcommand\be{\begin{equation}}
\newcommand\ba{\begin{eqnarray}}
\newcommand\ee{\end{equation}}
\newcommand\ea{\end{eqnarray}}
\newcommand{\bes}{\begin{subequations}}
\newcommand{\ees}{\end{subequations}}
\newcommand{\beqn}{\begin{eqnarray*}}
\newcommand{\eeqn}{\end{eqnarray*}}
\newcommand\bw{\begin{widetext}}
\newcommand\ew{\end{widetext}}

\newcommand{\nn}{\nonumber}

\newcommand{\SPA}{{\mbox{\tiny SPA}}}
\newcommand{\DFT}{{\mbox{\tiny DFT}}}
\newcommand{\p}{\partial}
\newcommand{\eff}{{\mbox{\tiny eff}}}
\newcommand{\inj}{{\mbox{\tiny inj}}}

\begin{document}

\title{Detection and parameter estimation of gravitational waves from compact binary inspirals with analytical double-precessing templates}

\author{Katerina Chatziioannou}
\affiliation{Department of Physics, Montana State University, Bozeman, Montana 59718, USA.}
\author{Neil Cornish}
\affiliation{Department of Physics, Montana State University, Bozeman, Montana 59718, USA.}
\author{Antoine Klein}
\affiliation{Department of Physics, Montana State University, Bozeman, Montana 59718, USA.}
\affiliation{Department of Physics and Astronomy, The University of Mississippi, University, MS 38677, USA.}
\author{Nicol\'as Yunes}
\affiliation{Department of Physics, Montana State University, Bozeman, Montana 59718, USA.}

\date{\today}

\begin{abstract}

We study the performance of various analytical frequency-domain templates for detection and parameter estimation of gravitational waves from spin-precessing, quasicircular, compact binary inspirals. We begin by assessing the extent to which nonspinning, spin-aligned, and the new (analytical, frequency-domain, small-spin) double-precessing frequency-domain templates can be used to detect signals from such systems. For effective, dimensionless spin values above $0.2$, the use of nonspinning or spin-aligned templates for detection purposes will result in a loss of up to $30\%$ of all events, while in the case of the double-precessing model, this never exceeds $6\%$. Moreover, even for signals from systems with small spins, nonspinning and spin-aligned templates introduce large biases in the extracted masses and spins. The use of a model that encodes spin-induced precession effects, such as the double-precessing model, improves the mass and spin extraction by up to an order of magnitude. The additional information encoded in the spin-orbit interaction is invaluable if one wishes to extract the maximum amount of information from gravitational wave signals.

\end{abstract}
\pacs{04.30.-w,04.80.Nn,04.30.Tv}

\maketitle

\section{Introduction}
\label{intro}

The detection of gravitational waves (GWs) will answer a plethora of important astrophysical questions about the population of compact objects in the nearby Universe. Second-generation, ground-based detectors, such as advanced~LIGO (aLIGO)~\cite{Abbott:2007kva,Harry:2010zz,ligo} and Advanced~Virgo (AdV)~\cite{Acernese:2007zze,virgonew,virgo}, are scheduled to resume operation in the next few years, with the first direct detections expected to follow shortly after. In preparation for all of this, the community is studying the most efficient ways of analyzing the forthcoming data, a nontrivial task for signals that are deeply buried in detector noise.

The most efficient way to extract and analyze such signals is through template filters. The latter are analytical or numerical models for the response of the detectors to impinging GWs. The templates are functions of a parameter vector $\vec{\theta}$ that characterizes the GW emitting system, and its position relative to the Earth. Parameter estimation consists of finding the components of $\vec{\theta}$ that best fit the signal, as well as their spread due to detector noise. Clearly, the efficiency of such an analysis is highly dependent on the accuracy of the template model itself~\cite{Cutler:2007mi,Sampson:2013jpa}.

One of the GW models we can construct most accurately represents waves emitted in the late-inspiral and merger of compact objects [neutron stars (NSs) and black holes (BHs)]. When these have masses less than 5 solar masses, the GWs emitted during the so-called \emph{inspiral} contribute the most to the signal-to-noise ratio (SNR), because of the frequency band (10-500)Hz in which the detector is most sensitive. During this phase, the binary components slowly orbit around each other, with orbital velocities $v \sim (0.05,0.4)c$, where $c$ is the speed of light. This allows one to construct a GW model perturbatively through the so-called post-Newtonian (PN) approximation, an expansion in $v/c$~\cite{Blanchet:2013haa}. The resulting PN model (and resummations thereof) has been shown to agree with purely numerical models up to the very last few orbits before plunge and merger~\cite{Buonanno:2006ui,Baker:2006ha,Pan:2007nw}.

But not all binary configurations can be accurately modeled through PN methods in a computationally efficient fashion. When the spin angular momentum of the binary components is misaligned with the orbital angular momentum, relativistic precession will induce GW modulations that are nontrivial to model~\cite{springerlink:10.1007/BF00756587,Bohe:2012mr}. This is why until recently most analytical modeling focused on nonspinning binaries (leading to nonspinning template models), binaries with spins aligned with the orbital angular momentum (spin-aligned template models), and binaries where only one component is spinning (simple precession template models). In the first two cases, the binary's orbital plane does not precess at all, while in the last case, the binary experiences simple precession, characterized by a single precession frequency~\cite{Apostolatos:1994mx}. 

A new \emph{purely analytic} way to construct generic double-precessing GWs has been recently proposed, the \emph{double-precessing model}, so named because the precession is characterized by two distinct frequencies. This model solves the precession equations through multiple-scale analysis~\cite{Klein:2013qda,bender}, a technique commonly employed in aeronautics, quantum field theory and more recently in relativity~\cite{Yunes:2005nn,Hinderer:2008dm}. Multiple-scale analysis is ideal to solve the orbital dynamics of inspiraling, precessing systems, because the latter have a natural separation of scales: the orbital time scale is much shorter than the precession time scale, which is much shorter than the radiation-reaction time scale. 

Two versions of the double-precessing model have been investigated so far, which are tailor made to describe different systems. The \emph{small-angle}, double-precessing model~\cite{Klein:2013qda} assumes the angle between the spin and the orbital angular momenta is small, while their magnitude is arbitrary. This model is well suited to BH binaries in a gaseous environment, since the latter tends to align the momenta~\cite{Bogdanovic:2007hp}. The \emph{small-spin}, double-precessing model~\cite{Chatziioannou:2013dza} assumes the magnitude of the spin angular momenta is small relative to the magnitude of the orbital angular momentum, while their orientation is arbitrary. This model is well suited to NS binaries, which are expected to have small spin magnitudes~\cite{Mandel:2009nx}.

Any template model, of course, is only as valuable to parameter estimation as it is accurate. In~\cite{Chatziioannou:2013dza}, we estimated the accuracy of the small-spin, double-precessing model relative to numerically constructed PN templates. The latter were obtained by numerically solving the Taylor-expanded PN precession equations, and then computing the discrete Fourier transform of the resulting, time-domain response function. The comparison between the analytic and the numerical PN models was carried out by calculating the so-called \emph{faithfulness} [see Eq.~\eqref{faithfulness-def}]: the normalized, noise-weighted inner product [see Eq.~\eqref{innerprod}] between a model and the signal, \emph{without} maximization over template parameters. The faithfulness is a good measure of the accuracy of the analytical template to recover the numerical PN model and estimate the latter's parameters. This measure was found to be above $98\%$ for NS binaries with dimensionless spin parameters up to $\chi_{A} = 0.2$, where $\chi_{A} \equiv S_{A}/m_{A}^2$ with $S_{A}$ and $m_{A}$ the magnitude of the spin angular momentum and the mass of the $A{\rm th}$ component respectively~\cite{Chatziioannou:2013dza}. 

Having established the accuracy of the analytical, small-spin, double-precessing model (from now on, we will refer to it as just double-precessing) relative to a purely numerical PN model, we now wish to study how good the former is at detecting and estimating the parameters of signals in noise. One expects that the double-precessing model should be able to recover more information from precessing signals, because it can capture the amplitude and phase modulations induced by precession, and thus, break degeneracies that are present in the absence of precession. We find that this is indeed the case: the precessing model breaks degeneracies between the mass ratio and the spin magnitudes~\cite{Cutler:1994ys}, allowing for a much better estimation of both quantities, by up to an order of magnitude. The improvement in parameter estimation is such that the precessing model can distinguish between NSs and BHs in the mass gap, \emph{even for nonspinning signals}~\cite{Chatziioannou:2014coa}. This result is in contrast to the conclusions one would arrive at if using spin-aligned templates that lack precession effects~\cite{Hannam:2013uu}. 

The idea that spin precession can significantly improve parameter extraction is by no means new. Vecchio~\cite{Vecchio:2003tn} was the first to show that spin-precession effects improve parameter extraction in the context of LISA sources. The restricted 1.5PN simple-precession model he considered~\cite{Apostolatos:1994mx} was later extended to 2PN order through numerical PN waveforms by Lang and Hughes~\cite{Lang:1900bz}, who reached similar conclusions. Klein et al.~\cite{PhysRevD.80.064027} included higher harmonics and showed that parameter extraction was further improved. Concerns that binaries in gas rich environments tend to have partially aligned spins, prompted Lang et al.~\cite{Lang:2011je} to study partially aligned models; they found that restricting precession degrades parameter extraction significantly, but the inclusion of higher harmonics improves extrinsic parameter extraction again. A similar result was recently found by O'Shaughnessy et al.~\cite{O'Shaughnessy:2014dka}. In another recent paper, Vitale et al.~\cite{Vitale:2014mka} performed an extensive search of the parameter space and found that parameter extraction is improved when precessional effects are maximized, i.e.~when the binary is observed edge on. The results of this paper, and those of~\cite{Chatziioannou:2014coa}, verify the above results and further demonstrate that the more accurate double-precessing model improves detection rates and parameter estimation for NS binaries so much so that it enables distinguishing between NSs and BHs and measure NS spins. The above comparison excludes the numerical PN templates, since they are slower by about a factor of $10^2$ or more than the double-precessing model~\cite{kyc}, a fact that makes them prohibitive for parameter estimation studies.

We here establish and explain these results in more detail by analyzing the performance of the nonspinning, the spin-aligned and the double-precessing models in detection and parameter estimation. Regarding detection, we study the efficiency of these templates at extracting a numerical PN model of GWs emitted by generically precessing, spinning binaries with arbitrary spin magnitudes. We address this by calculating the so-called \emph{fitting factor} [see Eq.~\eqref{ff-def}]: the normalized, noise-weighted inner product [see Eq.~\eqref{innerprod}] between a model and the signal, maximized over all template parameters. Such a measure is ideal to estimate how good a model is at recovering as much of the signal as possible at the expense of distorting the recovered parameters. This measure is above the nominal $98\%$ threshold, corresponding to a $6\%$ drop in detection rate, when using the spin-aligned and the small-spin, double-precessing templates for \emph{all} NS binaries with astrophysically realistic spins~\cite{Mandel:2009nx}. In the spin-aligned model, however, this large fitting factor comes at the expense of large biases in the extracted masses and spins. Binary BHs can have much larger spin magnitudes than NSs, and thus, the nonspinning and the spin-aligned models reach fitting factors above $98\%$ only for $\chi_{A} < 0.4$. The double-precessing model reaches fitting factors above this threshold for all $\chi_{A} < 1$, at the expense of large parameter biases.

We then consider the efficiency of these templates in parameter estimation, focusing on spin detectability and the accuracy in parameter extraction. In particular, we study what SNR and what injected spin parameter allows one to claim that a NS binary signal was produced by spinning NSs. If one can claim the signal corresponds to such a spinning binary, one can then address how well their spin magnitudes can be measured, again as a function of SNR and injected spin parameter. We tackle these questions in a Bayesian framework~\cite{Cornish:2007ifz,Littenberg:2009bm,Cornish:2011ys,Sampson:2013lpa,Sampson:2013jpa}, where we inject a small-spin, double-precessing signal and search for it through Markov-Chain Monte Carlo (MCMC) methods with either a spin-aligned template or a small-spin, double-precessing template. Such MCMC methods allow us to not only find the best fit parameters $\vec{\theta}_{\rm best}$, but also to construct their posterior probability distribution, as well as to determine which template model is best supported by the data.

One may be concerned that the small-spin, double-precessing template should not be used to estimate the statistical accuracy with which parameters can be inferred or alternative models distinguished, given that for higher spin values there will be systematic bias when recovering the true GWs we expect from nature. However, as we show in Appendix~\ref{app-errors}, systematic errors and statistical errors are independent for small model deviations, and the statistical errors found by using a waveform family that is close to the waveforms are nearly identical. Thus, the analytic double-precessing model can be used for reliable Bayesian inference and model selection.

The first parameter estimation question we tackle is that of spin detectability, which is a model selection problem~\cite{Gossan:2011ha,Li:2011cg,DelPozzo:2011pg,Aasi:2013jjl}: given a signal, one wishes to determine which of two competing models (``the signal was produced by a spinning binary'' versus ``the signal was produced by a nonspinning binary'') is best supported by the data. We address this problem by calculating the~\emph{Bayes factor} (BF), which provides an estimate of how well a model fits the data compared to another model. Since we are dealing with \emph{nested models} (models which reduce to each other when a subset of their parameters $\vec{\theta}$ acquire certain values), the BF can be calculated through the \emph{Savage-Dickey density ratio}~\cite{1995}: the ratio of the prior to the posterior evaluated at vanishing spins. We find that the data prefer the small-spin, double-precessing model over the nonspinning model at dimensionless spin magnitudes larger than roughly $0.02$ for SNR 10 with aLIGO~\cite{Chatziioannou:2014coa} and $0.01$ for SNR 30 with LIGO3~\cite{adhikari}. On the other hand, use of the spin-aligned model increases the spin detection threshold to roughly $0.05$ and $0.02$ respectively. 

The second parameter estimation question we address is that of accuracy in parameter extraction. Given a small-spin double-precessing signal, we determine the best-fit parameters and their 1-$\sigma$ confidence region (the smallest area in parameter space that contains $68\%$ of the posterior weight) for either a spin-aligned or a double-precessing model. We find that the small-spin, double-precessing templates can measure masses and spins roughly 1 order of magnitude better than spin-aligned templates. This is because even a small amount
of precession is sufficient to greatly deteriorate the likelihood of a double-precessing template, while a spin-aligned template cannot access this extra structure. This structure breaks degeneracies between the mass ratio and the spin magnitudes, allowing for a better measurement of both quantities. We show that these result are insensitive to the specific choice of spin priors: uniform over spin magnitudes and uniform over spin orientations on a 2-sphere. The improvement in parameter estimation is so dramatic that one should be able to distinguish between NS binaries and BH binaries purely from the detection of GWs during the inspiral phase.   

The remainder of the paper explains and expands the results described above in more detail. 
In Sec.~\ref{model}, we present the waveform models we use. 
In Sec.~\ref{ff}, we tackle the issue of detectability.
In Sec.~\ref{mcmc}, we study parameter estimation.
In Sec.~\ref{conclusions}, we conclude and point to future research.
Throughout the paper we use units where $G=c=1$.

\section{Waveform Models}
\label{model}
We consider BHBH binaries and NSNS binaries with masses $m_1$ and $m_2$ (where $m_1 \ge m_2$) and spin angular momentum magnitudes $S_1$ and $S_2$ respectively in adiabatically evolving, quasicircular orbits in the inspiral phase. GWs emitted from such a system induce a signal on ground-based detectors described by the parameter vector 
\begin{align}
\vec{\theta} = &({\cal{M}}, m, \cos{\theta_N}, \phi_N, D_L, \cos{\theta_L}, \phi_L, t_c, \phi_c, \nn \\
& \cos{\theta_1}, \phi_1, \chi_1, \cos{\theta_2}, \phi_2, \chi_2 ),
\end{align}
where ${\cal{M}} = (m_1 m_2)^{3/5}/(m_1+m_2)^{1/5}$ is the chirp mass, $m = m_1 + m_2$ is the total mass, $\cos{\theta_N}$ and $\phi_N$ are sky location angles, $D_L$ is the luminosity distance, $\cos{\theta_L}$ and $\phi_L$ are angles that describe the direction of the initial orbital angular momentum vector,  $\cos{\theta_A}$ and $\phi_A$ are angles that describe the direction of the initial spin angular momentum vectors, with $\chi_A = S_A/M^2_A$ the dimensionless spin magnitude, for the $A{th}$ binary component. All angles are measured in a geocentric frame~\cite{AdvLIGO-frame}.

In this paper, we consider five different waveform models: one that is purely numerical; two versions of the analytical, small-spin, double-precessing model; one version of the analytical spin-aligned model; and one version of the analytical nonspinning model. When considering detection issues in Sec.~\ref{ff}, we use the numerical PN model as the signal and the other four models as templates. When considering parameter estimation issues in Sec.~\ref{mcmc}, we use one of the double-precessing models as the signal, and a subset of the other analytical models as the template. We describe each of these models below.

{\emph{Numerical PN model}}. This model is constructed by first solving the most accurate PN spin-precession equations numerically (see e.g.~\cite{Klein:2013qda,Chatziioannou:2013dza}), and then Fourier-transforming the numerical PN  time-domain response function through a discrete Fourier transform. We use this model as the signal when studying detection issues in Sec.~\ref{ff}, but we do not use it as a template due to its high computational cost. We stress again that this model is constructed by solving PN ordinary differential equations numerically, similarly to the SpinTaylorT4~\cite{Buonanno:2009zt,Aasi:2013jjl}, or effective-one-body models~\cite{Buonanno:1998gg,Damour:2001tu}. Therefore, we regard it as a numerical PN model, in contrast to the closed-form analytical models we describe below, and the full numerical relativity based models of~\cite{Pan:2007nw,Ajith:2007qp,Duez:2009yz,Santamaria:2010yb}. Sometimes in the literature this model is referred to as `semianaltical'.

{\emph{Double-precessing models}}~\cite{Chatziioannou:2013dza}. The precession equations are solved by separating the three intrinsic time scales of the problem: the \emph{orbital} time scale, which is much shorter than the \emph{precession} time scale, which is much shorter than the \emph{radiation-reaction} time scale. The resultant orbital precession equations are then expanded in $\chi_{A} \ll 1$ and in the ratio of the different time scales. Such a \emph{multiple-scale analysis}~\cite{bender} treatment results in an analytical solution for the temporal evolution of the orbital and the spin angular momenta, valid to first order in $\chi_{A}$ and in the ratio of the precession to the radiation-reaction time scale. This solution can then be used to construct a time-domain response function that is Fourier-transformed through the \emph{stationary-phase approximation} (SPA)~\cite{Droz:1999qx,Yunes:2009yz}. Two versions of such a waveform can be constructed:
\begin{itemize}
\item[(i)] {\emph{Full, double-precessing:}} both the Fourier amplitude and phase are kept to high PN order [see Eqs. (105), (106), (107) of~\cite{Chatziioannou:2013dza}.] 
\item[(ii)] {\emph{Restricted, double-precessing:}} the Fourier amplitude is kept only to leading PN order, while all known PN corrections are kept in the Fourier phase [see Eqs. (98), (99), (100) of~\cite{Chatziioannou:2013dza}.]
\end{itemize} 
Henceforth, a term is said to be of N PN order if it scales as $(v/c)^{2N}$ relative to the leading order term in the expression, where $v$ is the binary's orbital velocity, and recall that $c$ is the speed of light.

{\emph{Restricted spin-aligned model}}. This waveform is constructed by assuming the spin angular momenta are \emph{exactly} aligned with the orbital angular momentum. Such an alignment prevents the system from precessing, thus rendering the spin-precession equations simple to solve~\cite{Blanchet:2013haa}. One then solves the evolution equation for the orbital frequency and phase through a PN expansion, which allows the construction of a time-domain response function. The latter is Fourier-transformed through the SPA. We here consider a restricted model, where only the leading PN order term is kept in the Fourier amplitude, while the Fourier phase is kept to $3.5$PN order\footnote{Terms beyond 3.5PN order are not completely known. Yet, we artificially extend the series to 8PN order, as explained in~\cite{Chatziioannou:2013dza}.}.The performance of these templates has been studied in numerous papers~\cite{Berti:2004bd,Agathos:2013upa,Aasi:2013jjl,Hannam:2013uu}; most of them conclude that, even though spin-aligned templates might be good enough for detection of NSNS binaries, they lead to large biases when used in parameter estimation. 

{\emph{Restricted nonspinning model}}. This waveform is derived assuming that the binary components have no spin angular momenta. The temporal evolution of the orbital frequency is obtained analytically through a PN expansion, which is then used to construct a time-domain response. The latter is Fourier transformed through the SPA. We here focus on a restricted version of these waveforms, where we keep the Fourier amplitude to leading PN order, but the Fourier phase is kept to $3.5$PN order. Such waveforms have been studied extensively in the literature~\cite{Arun:2004hn,Luna:2006gw,Ajith:2009fz,Rodriguez:2013oaa,O'Shaughnessy:2013vma,Veitch:2012df,Mikoczi:2012qy,Shah:2012vc}, mainly as detection templates, despite their inherent inability to measure spins. 

The waveforms described above are not the only ones that have been studied for detection and parameter estimation. A particularly interesting model has been constructed assuming one of the binary components has vanishing spin angular momentum~\cite{Apostolatos:1994mx,Vecchio:2003tn,vanderSluys:2008qx,Arun:2008zn,vanderSluys:2009bf,O'Shaughnessy:2014dka}. When this is the case, the nonvanishing spin evolves according to simple precession, allowing for a simple solution to the spin-precession equations. In this paper, we do not use this waveform model, since we consider BHBH binaries or NSNS binaries, which are likely to both have nonvanishing spin-angular momenta, albeit of different magnitude. The simple precession model would be useful for studying BHNS systems and it could be systematically improved through the formalism of~\cite{Klein:2013qda}.

When studying parameter estimation in Sec.~\ref{mcmc}, we will be in part interested in the errors associated with the extraction of parameters. There are two main types of errors in parameter recovery: \emph{systematic errors} and \emph{statistical errors}. Systematic errors are associated with a shift in the peak of the posterior distribution of the recovered parameter away from the injected value; such errors can be produced by inaccuracies in the template model. Statistical errors are associated with the inherent width of the posterior distribution; such an error is produced by the signals possessing a finite SNR. As we show explicitly in Appendix~\ref{app-errors}, systematic and statistical errors are independent to first order in the inaccuracies of the model, and thus, we will study them separately:  in Sec.~\ref{ff} we study the former through a fitting factor analysis; in Sec.~\ref{mcmc} we study the latter by investigating the shape of the posterior distribution surface around its maximum. For this reason, in Sec.~\ref{mcmc} we inject a signal created by the double-precessing model: we isolate the statistical errors by minimizing the systematic ones.

When studying statistical errors, not all analytical models will be used as templates. The nonspinning model is inappropriate for parameter estimation of spinning systems, and thus, it will not be used as a template in Sec.~\ref{mcmc}. The full double-precessing model is much more computationally expensive to evaluate, yet it is almost indistinguishable from the restricted double-precessing model. For these reasons, we will use only the restricted double-precessing and the spin-aligned model as templates in Sec.~\ref{mcmc}.
 
When studying systematic errors, we will randomize over all model parameters, while when focusing on statistical errors, we will select a few characteristic systems. By doing so, we isolate the effects of SNR, injected spin parameter and detector (aLIGO or LIGO 3) on parameter recovery. The three systems we work with are characterized by the parameters  in Table~\ref{systems}.

\begin{table}
\begin{centering}
\begin{tabular}{cccccccccc}
\hline
\hline
\noalign{\smallskip}
 {}   &&  $m_1$ &  $m_2$ & $\chi_1$ & $\chi_2$ & $\cos{\theta_N}$ & $\phi_N$ & $\iota$ & $\kappa$  \\
\hline
\noalign{\smallskip}
1 && 1.43 & 1.23 & *   &  * & -0.11 & 3.71 & $63^{\circ}$ & $30^{\circ}$\\
2 && 1.43 & 1.23 & 0.04 & 0.04 & -0.11 & 3.71 & $63^{\circ}$ & *\\
3 && * & * & 0.04  & 0.04 & -0.11 & 3.71 & $63^{\circ}$ & $30^{\circ}$\\
\noalign{\smallskip}
\hline
\hline
\end{tabular}
\end{centering}
\caption{Summary of the systems used in the parameter estimation analysis of Sec.~\ref{mcmc}. The masses are in units of solar masses, $\iota$ is the angle between the orbital angular momentum and the line of sight at GW frequency $70$Hz, and $\kappa$ is the opening angle between the orbital angular momentum and the total spin angular momentum again at $70$Hz. The asterisk denotes the parameters that are varied.}
\label{systems}
\end{table}
%

\section{Detection}
\label{ff}

In GW astronomy, there are two measures that estimate the extent to which two models are similar to each other: the \emph{faithfulness} and the \emph{fitting factor}~\cite{Damour:1997ub}. In our case, the signal will always be the numerical PN model, while the template will be the analytical models described in Sec.~\ref{model}. Both measures depend on the noise-weighted inner product between two models for the response function, $h_{1}$ and $h_{2}$:
\begin{align}
\left(h_{1}\left|\right.h_{2}\right) &\equiv 4 \Re\int_{f_{\min}}^{f_{\max}}\frac{\tilde{h}_1(f) \tilde{h}_2^*(f)}{S_{n}(f)}\;df\,,
\label{innerprod}
\end{align}
where the overhead tilde stands for the Fourier transform, $\Re[\cdot]$ is the real part operator, $(f_{\min},f_{\max})$ are the limits of integration, and $S_{n}(f)$ is the detector's spectral noise density; we here use the high-power, zero-detuned $S_{n}(f)$ of aLIGO~\cite{AdvLIGO-noise}. 

The faithfulness is a measure of how good a template is at recovering a signal with the \emph{same} parameters, and thus, how efficient the model is at parameter recovery. It is defined as
\begin{align}
F_{h_1,h_2} &\equiv \text{max}_{t_c, \phi_c} \frac{\left(h_{1}\left|\right.h_{2}\right)}{\sqrt{\left(h_{1}\left|\right.h_{1}\right)
\left(h_{2}\left|\right.h_{2}\right)}},\label{faithfulness-def}
\end{align}
where the inner product is maximized only over the time of coalescence $t_c$ and the phase of coalescence $\phi_c$. 

The fitting factor is a measure of how good a template is at recovering a signal regardless of biasing parameter recovery. It is defined through
\begin{align}
FF_{h_1,h_2} &\equiv \text{max}_{\vec{\theta}} \, \frac{\left(h_{1}\left|\right.h_{2}\right)}{\sqrt{\left(h_{1}\left|\right.h_{1}\right)
\left(h_{2}\left|\right.h_{2}\right)}}\,,
\label{ff-def}
\end{align}
where the inner product is maximized over all parameters. In general, the highest $FF$ is achieved between models $h_{1}$ and $h_{2}$ that have \emph{different} parameters. 

In order to obtain reliable estimates for these two measures that are independent of the specific system considered, we create a random distribution of systems through Monte Carlo (MC) methods. The mass distribution is chosen to be flat in log space, with boundaries chosen depending on the class of system considered: for NS binaries, we choose the range $[1,2.5] M_{\odot}$, while for BH binaries, we choose the range $[5,10] M_{\odot}$. All vector directions are chosen uniformly on the sphere. We present our results as a function of the symmetric dimensionless spin parameter $\chi_s \equiv (\chi_1+\chi_2)/2$ since it gives a measure of how applicable the small-spin approximation of~\cite{Chatziioannou:2013dza} is for the particular system studied. Below we study BH binaries separately from NS binaries, since the latter are expected to have comparable masses and small spin magnitudes, while the former are not.

\subsection{NSNS Binaries}
\label{NSNS}

NS binaries that enter the sensitivity band of ground-based detectors are not expected to have large spin magnitudes. This is because although NSs can be spun up by accretion, they spin down due to magnetic breaking. By the time they have spiraled into each other sufficiently to be emitting GWs detectable by ground-based detectors, their spin magnitudes are not expected to exceed $\chi_{A} = 0.2$~\cite{Mandel:2009nx}. This fact makes NSNS binaries an ideal candidate for the small-spin, double-precessing model of~\cite{Chatziioannou:2013dza}.  

\begin{figure*}[t]
\begin{center}
\includegraphics[width=\columnwidth,clip=true]{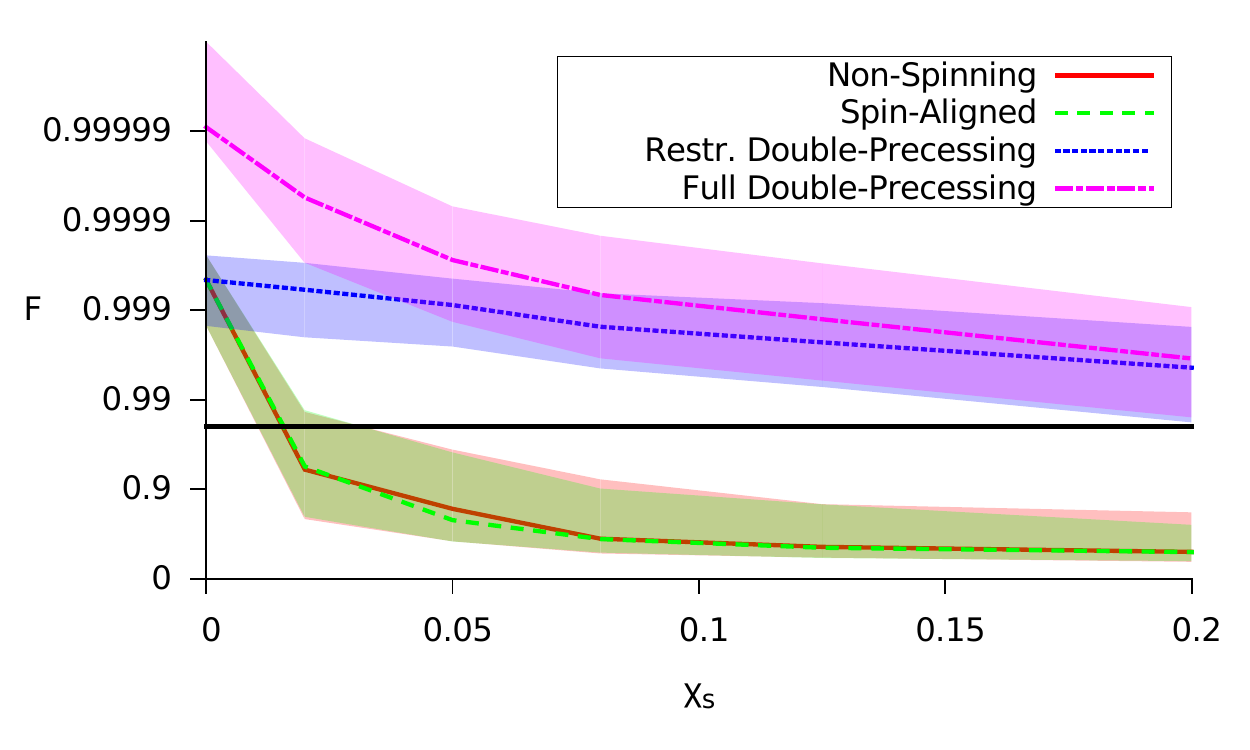}
\includegraphics[width=\columnwidth,clip=true]{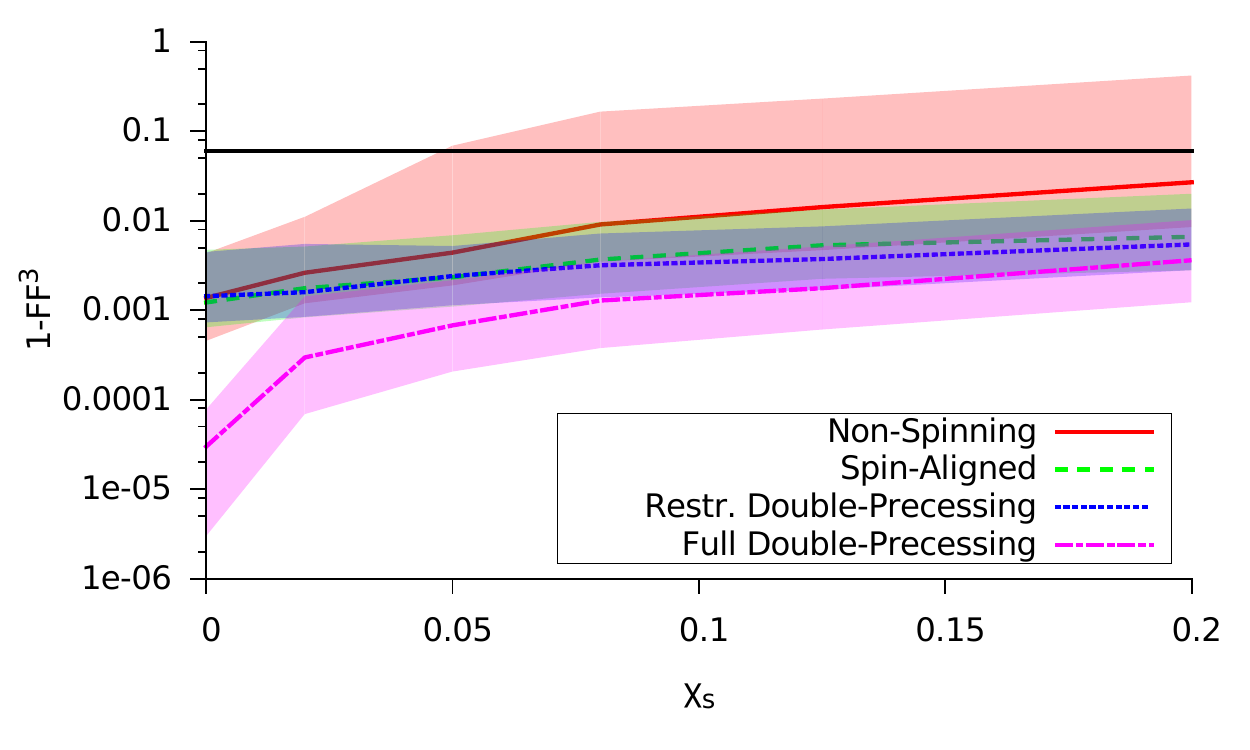}
\caption{\label{fig:matchandff_NSNS} (Color Online) Median faithfulness (left panel) and median drop in detection rates (right panel) between a numerical PN waveform and a full double-precessing waveform (magenta dot-dashed line), a restricted double-precessing waveform (blue dotted line), a restricted spin-aligned waveform (green dashed line), and a restricted nonspinning waveform (red sold line) for NSNS binaries as a function of the symmetric spin. The shaded areas give the 1-$\sigma$ confidence regions and the black solid line represents the $98\%$ threshold. In the case of detection rates this threshold corresponds to the loss of $6\%$ of all events.}
\end{center}
\end{figure*}

Figure~\ref{fig:matchandff_NSNS} shows the faithfulness and (one minus the cube of) the fitting factor for NSNS binaries between the numerical PN model and all the analytic models (see Sec.~\ref{model} for a description), as a function of the symmetric dimensionless spin parameter. Since the recovered SNR of a source scales as the fitting factor, $1-\rm{FF}^3$ gives an estimate of the reduction of the volume accessible to the detectors due to model inaccuracies. In other words, when the fitting factor drops, the source needs to be closer to earth to give the same SNR value and be detectable. For this reason we interpret $1-\rm{FF}^3$ as the drop in overall expected detection rates of aLIGO/AdV, which are highly uncertain to begin with~\cite{Abadie:2010cf}.

Each point in $\chi_{s}$ is computed by averaging over 2000 random systems (600 for the full double-precessing model due to computational restrictions) with masses in $[1,2.5] M_{\odot}$. The lower limit of integration is $f_{\min}=10$Hz, the frequency at which GWs enter the aLIGO band. The upper limit of integration is $f_{\max}=400$Hz, while the system is still in the inspiral phase, in order to avoid finite size effects that enter above this frequency~\cite{Read:2009yp,Hinderer:2009ca,Markakis:2010mp}. 

Three primary conclusions can be drawn from these plots. First, the faithfulness stays above the nominal $98\%$ threshold when using the double-precessing models for all spins considered, while it drops below this threshold for the nonspinning and spin-aligned system above $\chi_{s} = 0.02$. This indicates that only the double-precessing models can be considered as reliable parameter estimation templates. Second, the fitting factor is above the $98\%$ threshold, corresponding to a loss of event rate smaller than $6\%$, for all models. As expected from previous results, the nonspinning and the spin-aligned models can serve as detection templates for slowly spinning systems, like NS binaries. Third, we find similar fitting factors when using the restricted and the full double-precessing models. This implies that the restricted model is sufficient for parameter estimation studies.

\begin{figure*}[t]
\begin{center}
\includegraphics[width=\columnwidth,clip=true]{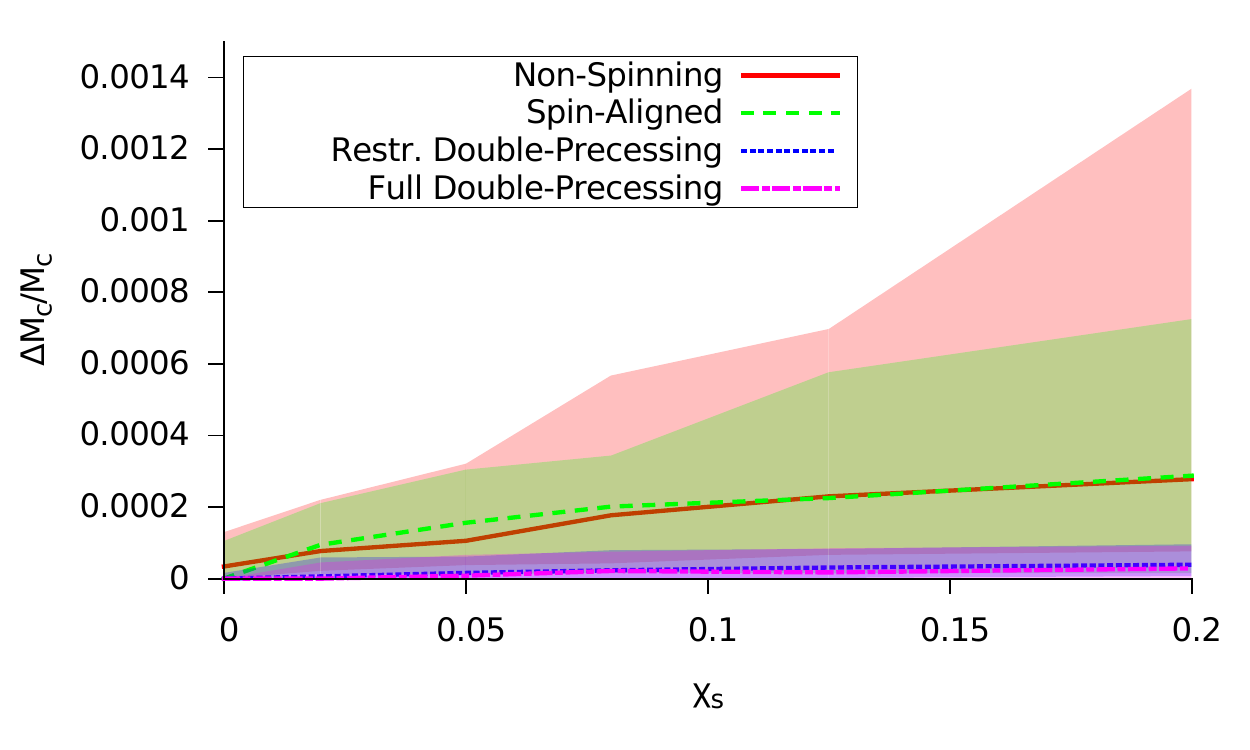} \quad
\includegraphics[width=\columnwidth,clip=true]{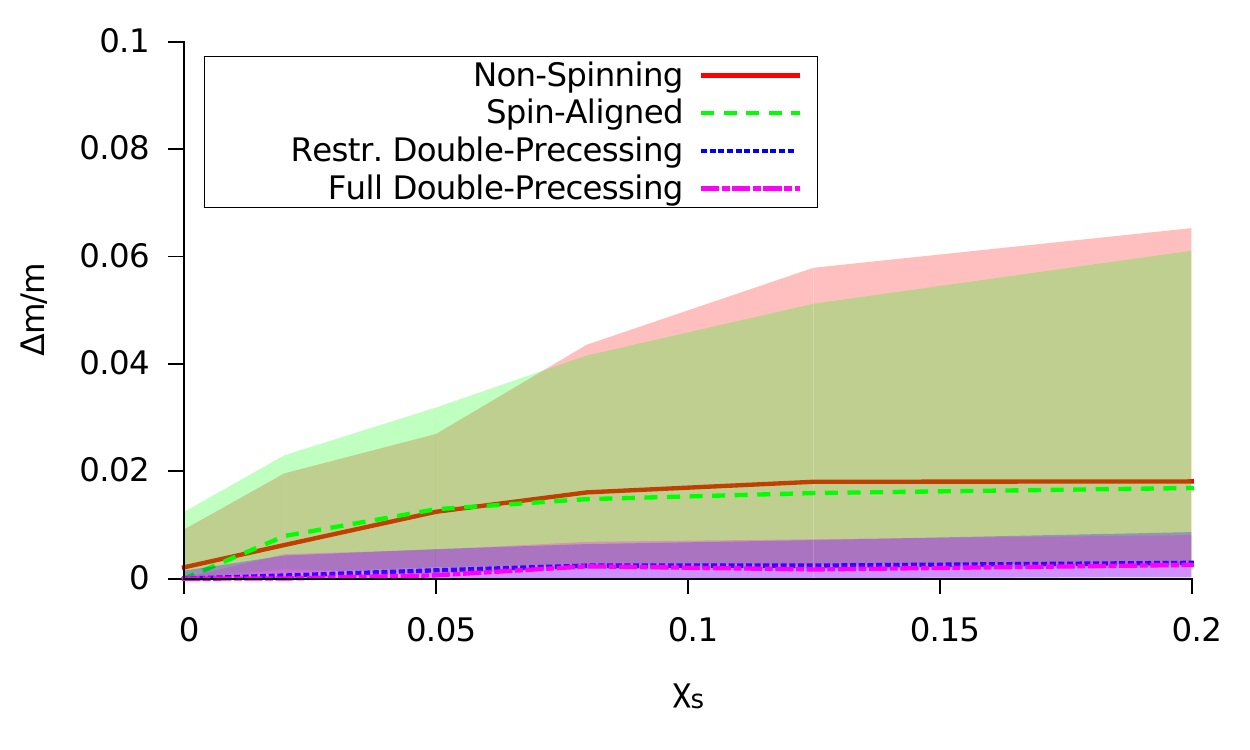} \\
\includegraphics[width=\columnwidth,clip=true]{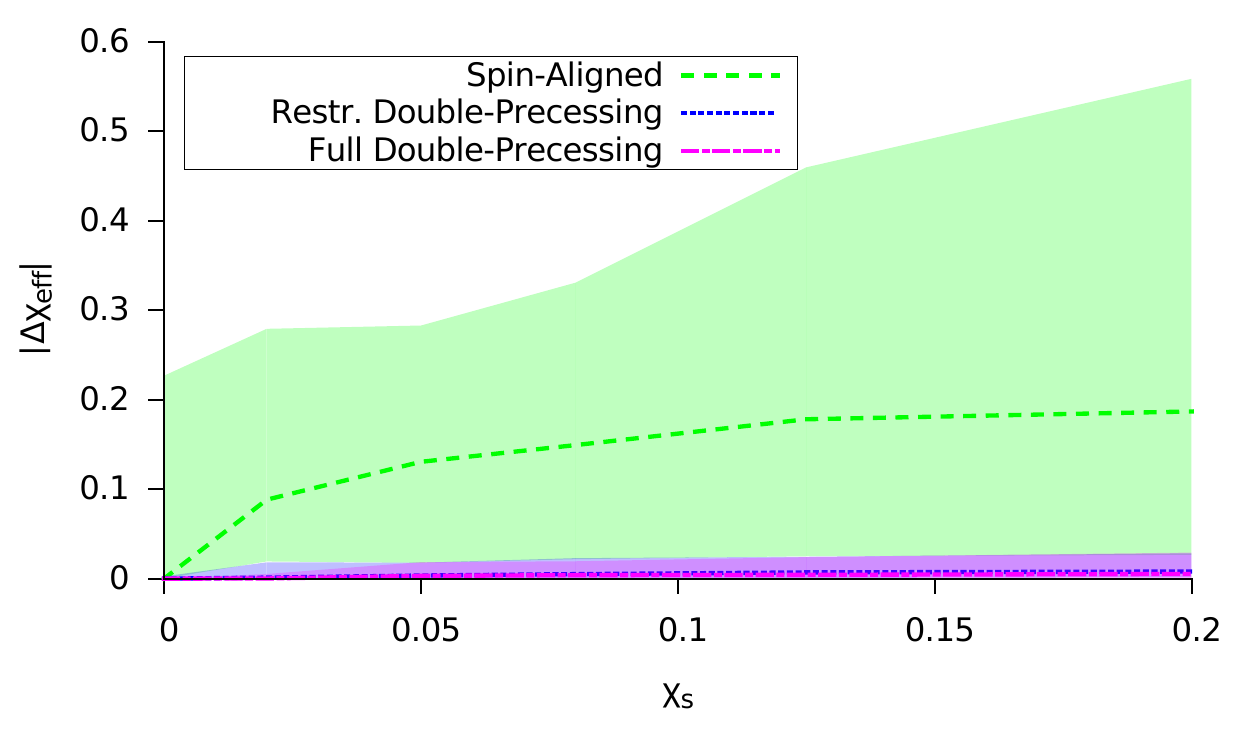} 
\caption{\label{fig:biases_NSNS} (Color Online) Median parameter bias for the chirp mass (top left), the total mass (top right), and the absolute value of the dimensionless effective spin parameter (bottom) for full double-precessing waveforms (magenta dot-dashed line), restricted double-precessing waveforms (blue dotted line), restricted spin-aligned waveforms (green dashed line), and restricted nonspinning waveforms (red sold line) for NSNS binaries as a function of the symmetric spin parameter. The shaded areas give the 1-$\sigma$ confidence regions. The bias from using the nonspinning, or spin-aligned templates is about an order of magnitude larger than the bias from the double-precessing templates. Also, the similar performance of the full and the restricted double-precessing templates makes the computationally less expensive restricted templates ideal for the parameter estimation studies of Sec.~\ref{mcmc}.}
\end{center}
\end{figure*}

Comparing the two panels of Fig.~\ref{fig:matchandff_NSNS} we see how the spin-aligned and nonspinning templates are able to distort their parameters to achieve a better overlap with the numerical PN model. In Fig.~\ref{fig:biases_NSNS} we plot the bias that such shifting induces on the chirp mass, the total mass, and the absolute value of the effective spin parameter [the symmetric spin combination projected onto the orbital angular moments; see Eq.~\eqref{effspin-def}]. Clearly, if the nonspinning or the spin-aligned waveforms are used for parameter estimation, the resulting parameter bias will be significant, and the systematic error will most likely dominate the total error.
%

\subsection{BHBH Binaries}
\label{BHBH}

Unlike NSs, there is no astrophysical reason to limit the spin magnitude of BHs (other than cosmic censorship, $\chi \leq 1$). One may thus expect the small-spin, double-precessing model of~\cite{Chatziioannou:2013dza} to perform badly when attempting to detect highly spinning signals. However, we find this not to be the case, due to the ability of the double-precessing model to shift its $15$ parameters in order to recover as much of the signal as possible .

\begin{figure*}[t]
\begin{center}
\includegraphics[width=\columnwidth,clip=true]{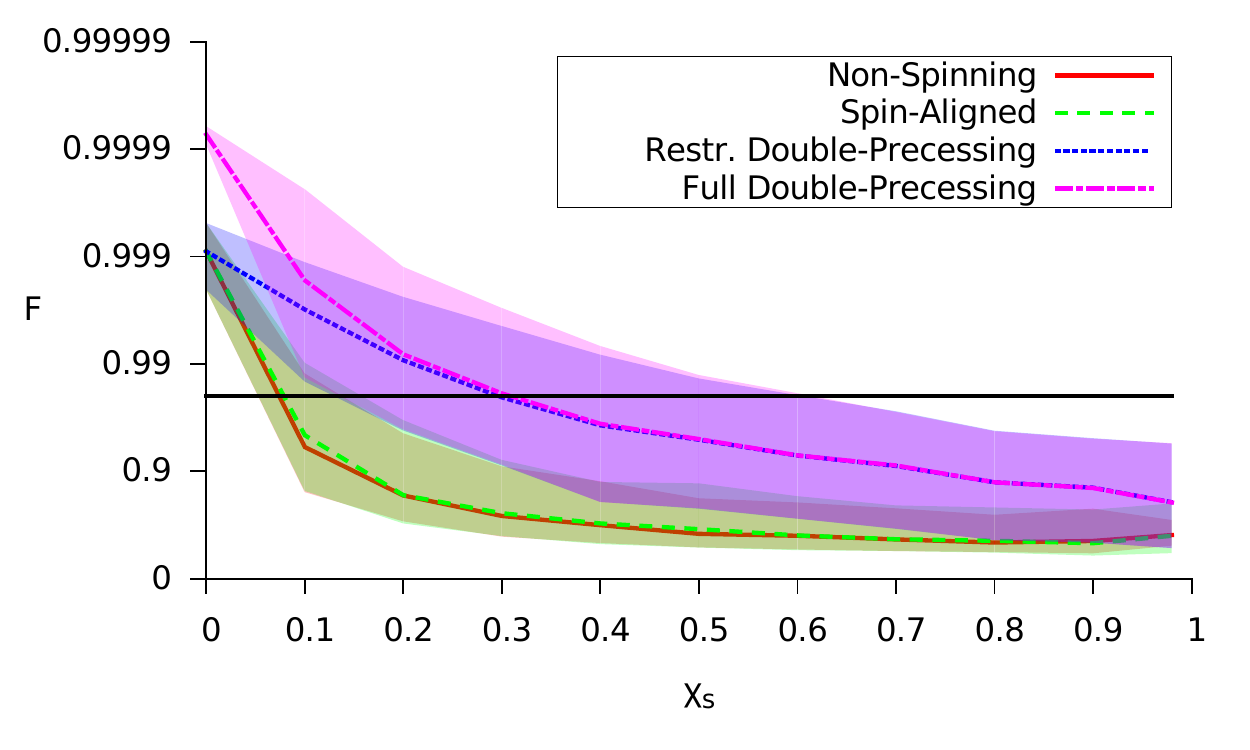}
\includegraphics[width=\columnwidth,clip=true]{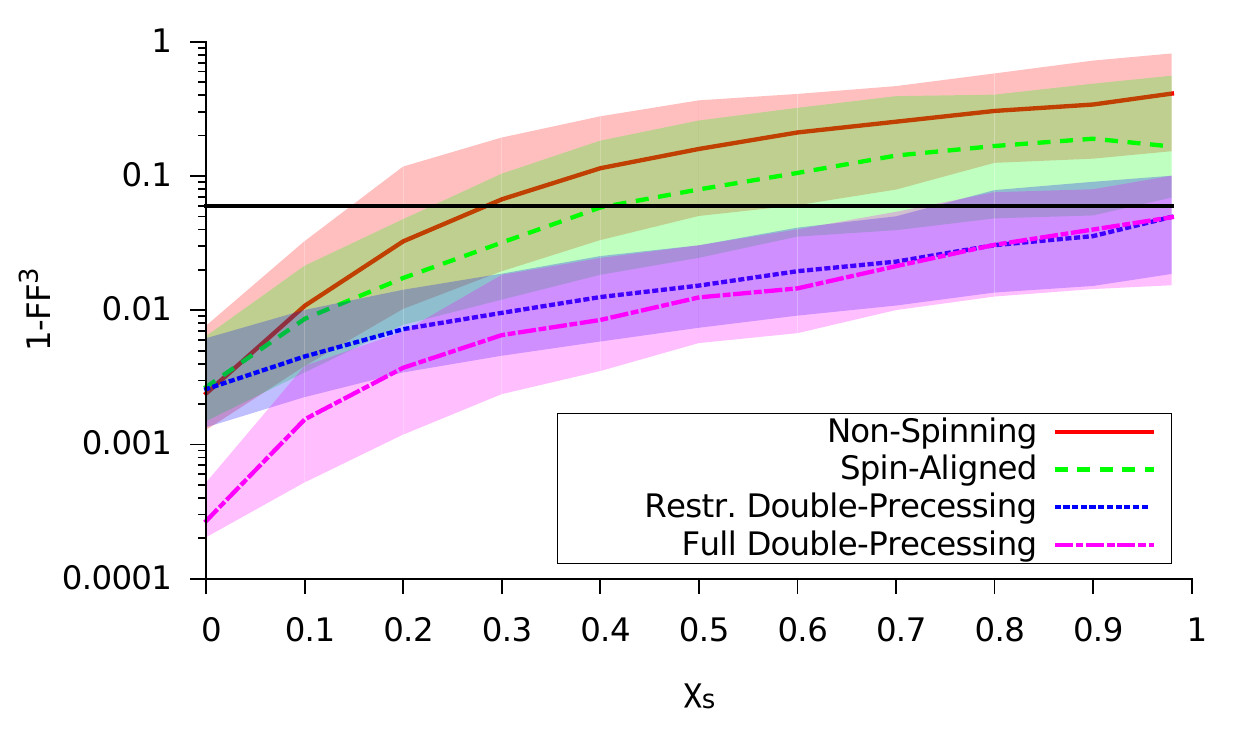}
\caption{\label{fig:matchandff_BHBH} (Color Online) Median faithfulness (left panel) and median drop in detection rates (right panel) between a numerical PN waveform and a full double-precessing waveform (magenta dot-dashed line), a restricted double-precessing waveform (blue dotted line), a restricted spin-aligned waveform (green dashed line), and a restricted nonspinning waveform (red sold line) for BHBH binaries as a function of the symmetric spin. The shaded areas give the 1-$\sigma$ confidence regions and the black solid line represents a value of $98\%$ (corresponding to a $6\%$ drop in detection rates). The use of nonspinning or spin-aligned templates for the detection of generically spinning BHBH binaries will result in a the loss of most highly spinning systems. On the other hand, both double-precessing models are capable of capturing most of the systems.}
\end{center}
\end{figure*}

Figure~\ref{fig:matchandff_BHBH} shows the faithfulness and (one minus the cube of) the fitting factor for BHBH binaries between the numerical PN model and all the analytic models as a function of $\chi_{s}$. Each point in $\chi_{s}$ is computed by averaging over 6000 random systems (1100 for the full double-precessing model due to computational restrictions) with masses in $[5,10] M_{\odot}$. The lower limit of integration is again set at $f_{\min}=10$Hz. However, since GWs emitted by BHs do not have any finite size effects, we extend the integration to the frequency corresponding to GWs emitted by a test particle at the innermost stable circular orbit (ISCO) of a Schwarzschild BH, i.e.~$f_{\max} = 6^{-3/2}/(\pi m)$. 

For spins larger than about $\chi_{s} = 0.3$ the faithfulness drops below $98\%$. Therefore, a first order expansion in the spins seems inadequate in capturing the strong precessional effects present in binaries with large spins. However, if the waveforms are allowed to adjust their parameters to fit the signal, they perform significantly better in detecting sources with large spin magnitudes. The right panel of Fig.~\ref{fig:matchandff_BHBH} shows the drop in detection rates for all analytical models. Now the double-precessing waveforms obtain overlaps greater than $98\%$ for all dimensionless spin magnitudes. The nonspinning and spin-aligned templates perform adequately for spins only up to $0.3$ and $0.4$ respectively. Clearly, the two double-precessing models are the only reliable detection templates for highly spin-precessing BHBH binaries of all models considered here.

\begin{figure*}[t]
\begin{center}
\includegraphics[width=\columnwidth,clip=true]{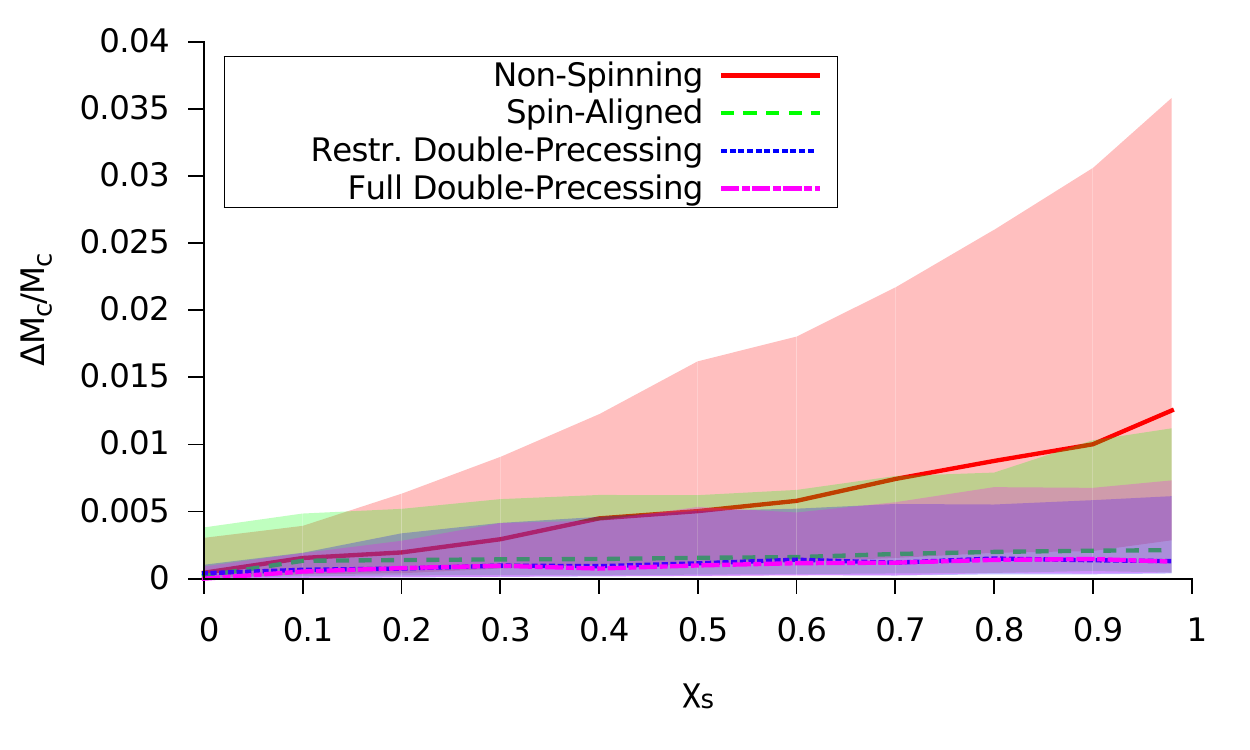} \quad
\includegraphics[width=\columnwidth,clip=true]{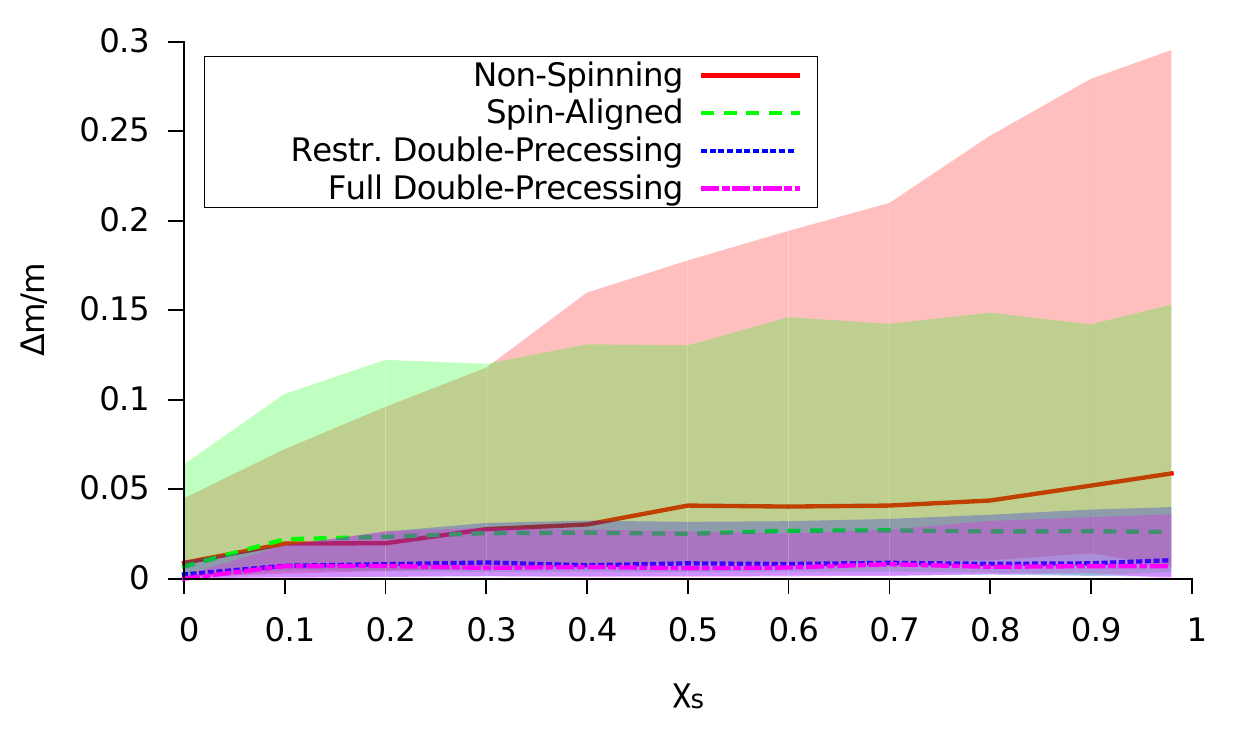} \\
\includegraphics[width=\columnwidth,clip=true]{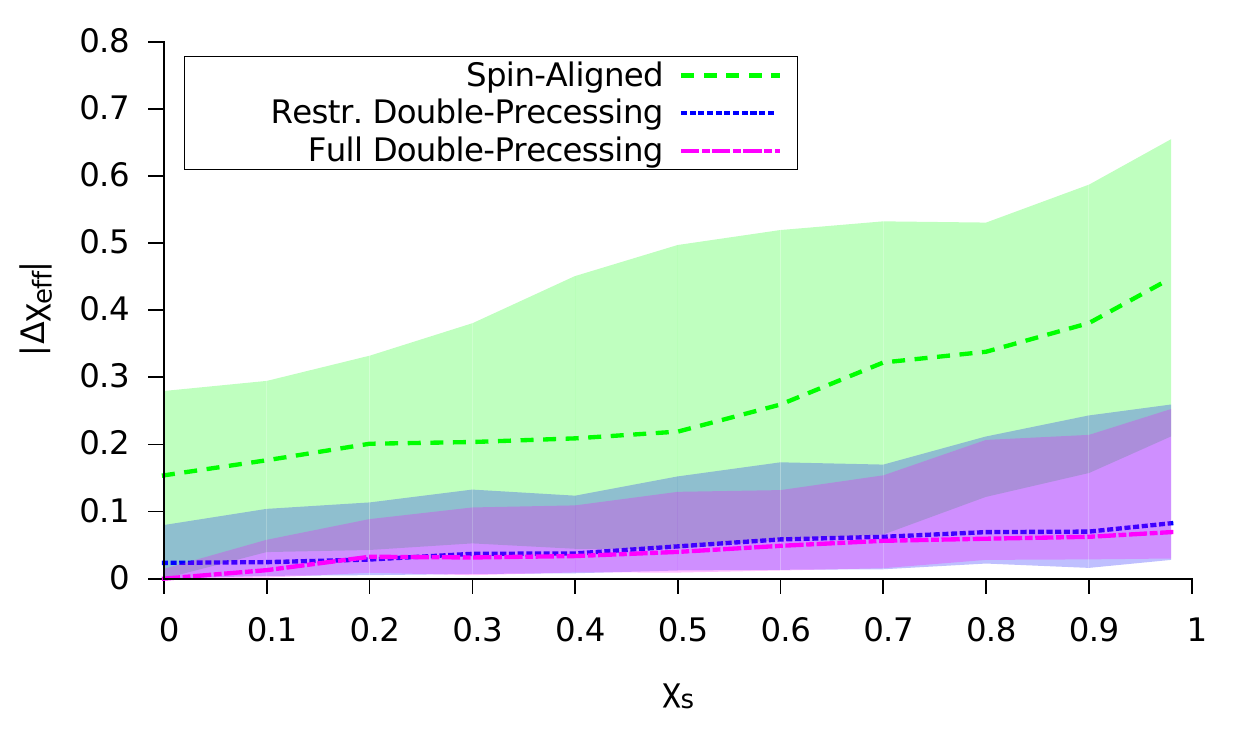} \quad
\caption{\label{fig:biases_BHBH} (Color Online) Median parameter bias for the chirp mass (top left), the total mass (top right), and $|\chi_{\eff}|$ (bottom) for full double-precessing waveforms (magenta dot-dashed line), restricted double-precessing waveforms (blue dotted line), restricted spin-aligned waveforms (green dashed line), and restricted nonspinning waveforms (red sold line) for BHBH binaries. The shaded areas give the 1-$\sigma$ confidence regions. All four models induce a significant amour of bias, even though the double-precessing ones perform much better.}
\end{center}
\end{figure*}

The large difference between faithfulness and fitting factor shows that even the double-precessing waveforms have to adjust their parameters significantly to achieve high overlaps with the numerical PN waveforms. Figure~\ref{fig:biases_BHBH} shows the bias in the chirp mass, the total mass, and the absolute value of the effective spin [Eq.~\eqref{effspin-def}]. The bias induced by using double-precessing templates is about an order of magnitude smaller than that incurred when using nonspinning or spin-aligned templates. Nonetheless, even the double-precessing templates induce a significant bias, making them unsuitable for parameter estimation of BHBH binaries~\cite{vanderSluys:2009bf,Aasi:2013jjl}. 

\subsection{Likelihood as a function of mass}
\label{LofM}

Maximizing the fitting factor reduces to maximizing the~\emph{likelihood} [the importance of which will become more evident in Sec.~\ref{mcmc}, where we carry out parameter estimation; see also Eq.~\eqref{likelihood-def}]. The efficiency of any maximization algorithm is highly dependent on our understanding of the behavior of the likelihood surface. A good understanding of this surface allows us to propose better jumps that, in turn, allow us to find the peak of this surface faster and overall explore it more efficiently (see Sec.~\ref{mcmc} for more details). It is, therefore, important to study how precession affects the likelihood surface.

\begin{figure*}[t]
\begin{center}
\includegraphics[width=\columnwidth,clip=true]{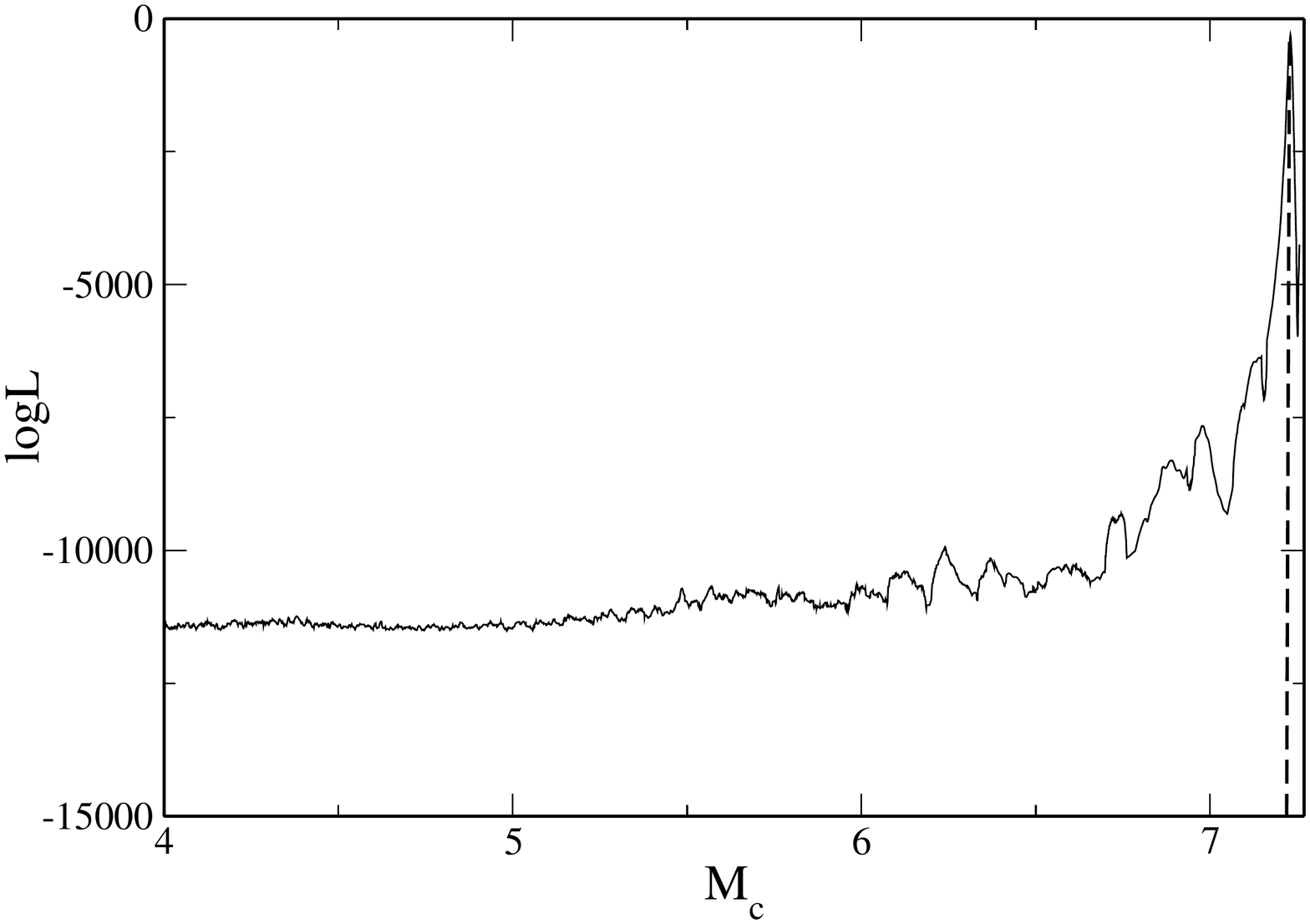}\quad
\includegraphics[width=\columnwidth,clip=true]{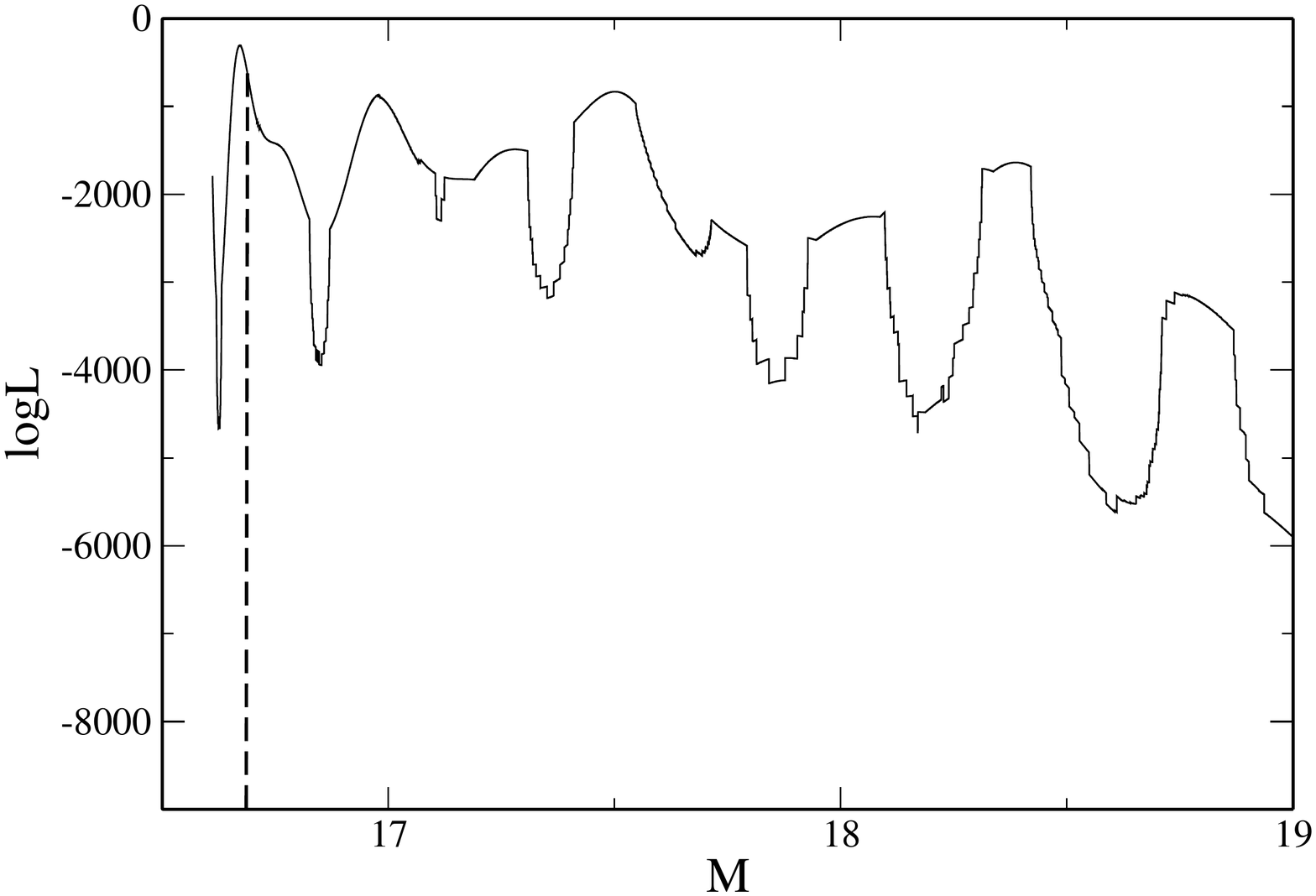}
\caption{\label{fig:logLofmasses} Log likelihood as a function of the chirp mass (left) and of the total mass (right) for a BHBH system. The vertical dashed lines correspond to the injected values ${\cal{M}}=7.23 M_{\odot}$ and $M= 16.7 M_{\odot}$.}
\end{center}
\end{figure*}

Figure~\ref{fig:logLofmasses} shows the log of the likelihood maximized over the time of coalescence, the phase of coalescence, and the luminosity distance as a function of the chirp mass (left panel) and as a function of the total mass (right panel) for a BHBH system with ${\cal{M}}=7.23 M_{\odot}$, $M= 16.8 M_{\odot}$, and $\chi_1=\chi_2=0.5$. The left panel shows a strong preference for the injected value; it is unlikely that a different chirp mass will be recovered. Indeed, when we studied parameter biases in Figs.~\ref{fig:biases_NSNS} and~\ref{fig:biases_BHBH}, we found that the chirp mass is biased by about $0.01\%$ for NSNS binaries and $0.3\%$ for BHBH binaries, depending on the spin of the injection.

On the other hand, the right panel of Fig.~\ref{fig:logLofmasses} shows a completely different dependence of the likelihood on the total mass. We see that the log of the likelihood presents a series of peaks with comparable heights. Therefore, as the other parameters in the model are varied from their injected value, one of the secondary peaks might become the primary one, resulting in a higher fitting factor. In fact, this is the case for the system presented here. The injected value for the total mass corresponds to the first peak around $M=16.8 M_{\odot}$. However, the recovered, or best fit, value for the total mass corresponds to the second peak at around $M=16.9 M_{\odot}$. By appropriately adjusting its parameters, the template model managed to find a better fit to the signal than by using the signal's parameters, resulting in a fractional systematic error of about $0.5\%$ for the recovered total mass. This is also verified by the top, right panel of Fig.~\ref{fig:biases_BHBH}. 

The series of peaks in log likelihood suggests that to map the likelihood surface sufficiently one should propose jumps between peaks, so that the Markov chains do not get stuck in a local maximum. To do so, we used the log likelihood as a function of the total mass maximized over $t_c, \phi_c$ and $D_L$ as an additional jump proposal, where all other parameters were held fixed, and the new total mass point was drawn from this distribution though rejection sampling. These jumps ensure that all peaks are explored adequately and the one with the maximum likelihood is selected.

An interesting consequence of the behavior of the likelihood surface is related to~\emph{theoretical bias}~\cite{Cutler:2007mi}. The latter is defined as the mismodeling error in parameter recovery induced by inaccuracies in the template model, e.g.~due to truncation of the PN series. One semianalytic estimate of this error can be obtained by modeling the likelihood surface as a single peak of finite width~\cite{Cutler:2007mi}: 
\be
\label{mis-modeling-error}
\Delta_{th}\theta^i = \left(\Gamma^{-1}(\theta_{tr})\right)^{ij}\left(\p_j h_{\SPA}(\theta_{tr})\left|\right. h_{\DFT}(\theta_{tr})-h_{\SPA}(\theta_{tr})\right),
\ee
where $h_{\DFT}$ is the true signal (in our case, the numerical PN model), $h_{\SPA}$ is the ``incorrect'' template that is used (in our case, any of the analytical models), and $ \left(\Gamma^{-1}\right)^{ij}$ is the inverse of the Fisher information matrix 
\be
\Gamma_{ij} =  \left(\p_i h \left|\right. \p_j h\right),
\ee
where $\p_i$ denotes differentiation with respect to the $i\rm{th}$ parameter. All quantities are evaluated at the injected parameters $\theta_{tr}$.

\begin{figure*}[t]
\begin{center}
\includegraphics[width=\columnwidth,clip=true]{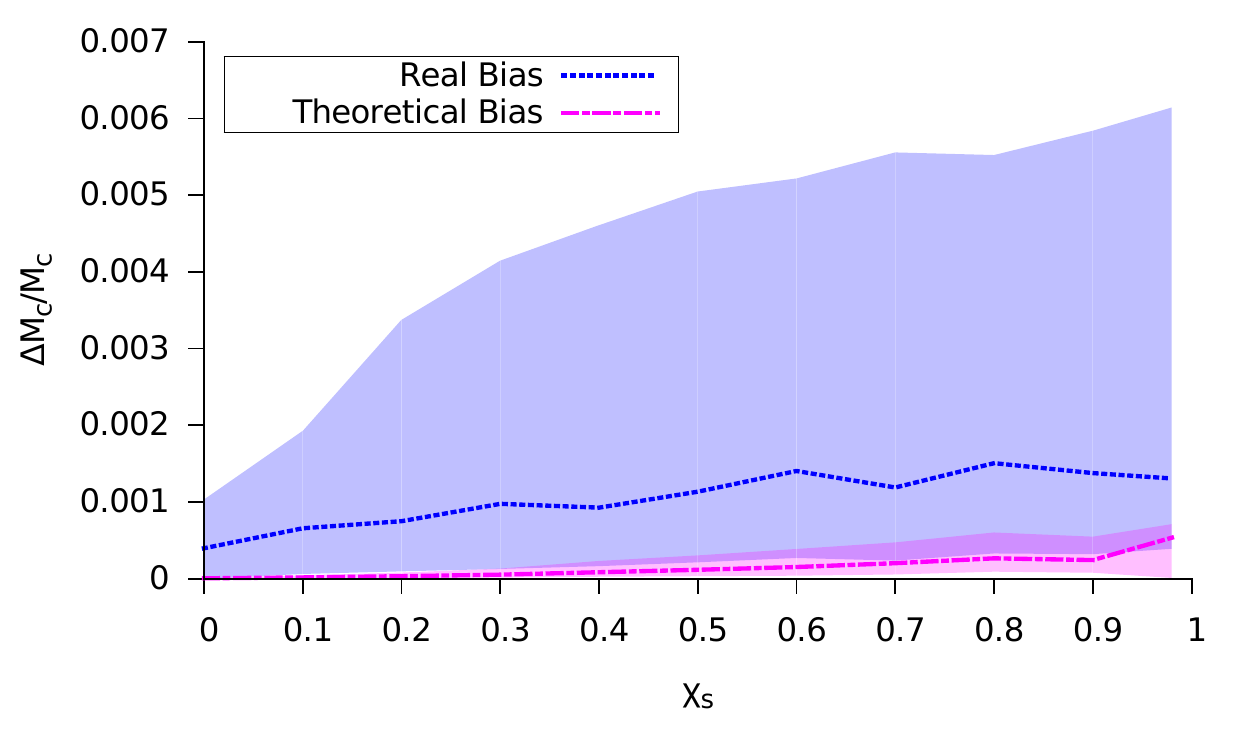}\quad
\includegraphics[width=\columnwidth,clip=true]{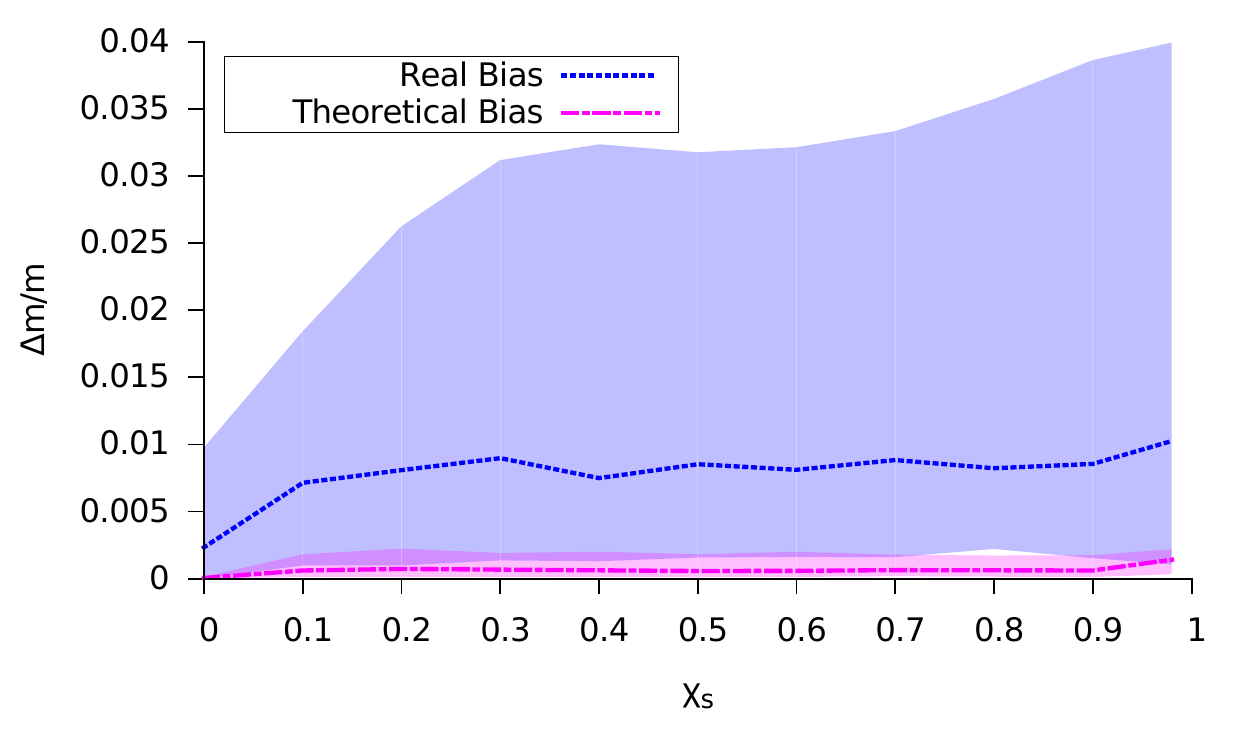}
\caption{\label{fig:TheoreticalvsReal_Bias_BHBH} (Color Online) Real bias (blue dotted line) and theoretical bias (magenta dot-dashed line) for the chirp mass (left panel) and the total mass (right panel) for a BHBH binary as a function of the injected spin. The shaded regions give the 1-$\sigma$ confidence intervals.}
\end{center}
\end{figure*}
This estimate of the mismodeling error due to theoretical bias is indeed approximately correct when the likelihood surface is single peaked, but it can grossly underestimate the biases when the surface is multipeaked. Figure~\ref{fig:TheoreticalvsReal_Bias_BHBH} shows the real error, as estimated from the posterior distribution, and the mismodeling error, as estimated with Eq.~\eqref{mis-modeling-error}, for the chirp mass (left panel) and the total mass (right panel)  as a function of the injected $\chi_{s}$, given a BHBH binary signal. For the total mass, Eq.~\eqref{mis-modeling-error} (roughly the width of the first peak) underestimates the true bias (roughly the distance between peaks) by an order of magnitude. As a further verification of this, we restricted the total mass range around the primary peak of the log likelihood and found that the real bias agreed with the theoretical bias.

\section{Parameter Estimation}
\label{mcmc}

Parameter estimation in the Bayesian framework results in an approximation to the posterior belief that a certain model with a parameter vector $\vec{\theta}$ describes the data $d$. The posterior distribution is calculated through Bayes' theorem
\be
p(\vec{\theta} | d) = \frac{p(d | \vec{\theta}) p(\vec{\theta})  }{p(d)},
\ee
where $p(\vec{\theta} | d)$ is the posterior belief, $ p(\vec{\theta})$ is the prior belief on the parameters, $p(d)$ is the evidence (here, an irrelevant normalization factor), and $p(d | \vec{\theta})$ is the~\emph{likelihood} that data $d$ were produced by a model with parameters $\vec{\theta}$. For the prior, we choose uniform distributions in the allowed region of parameter space. In GW studies, the likelihood is the noise model, which we here assume to be Gaussian and stationary
\be
p(d | \vec{\theta}) \sim \exp{\left[-\frac{1}{2} \left(s-h\left|\right.s-h\right)\right]}, \label{likelihood-def}
\ee
with $s$ the detectors' output and $h$ the template model. The $15-$dimensional posterior distribution is sampled though an MCMC algorithm. The posterior distribution for each parameter is then obtained by marginalizing over all other parameters.

Unlike in the previous section, where we were interested in maximizing the likelihood to recover the maximum overlap between the signal and the model, we are now interested in the likelihood surface itself. For that reason, as explained in Sec.~\ref{model}, we use the restricted double-precessing waveform as the injection and recover it with the spin-aligned model and the double-precessing model as templates. A very wide likelihood surface results in poor parameter extraction, while a peaked likelihood results in small errors for the recovered parameters.

Henceforth, we assume GW detections with the following three-detector network configurations: (i) two aLIGO detectors~\cite{Harry:2010zz} and one AdV~\cite{virgonew} detector with network SNRs of $10$ and $20$, and (ii) three detectors with the LIGO3 noise model~\cite{adhikari} and SNRs of $30$ and $60$. We concentrate on observations of the characteristic systems discussed in Sec~\ref{model} and described in Table~\ref{systems}. 

Furthermore, since we are dealing with NS binaries, we stop our analysis at a GW frequency of $400$Hz in order to avoid finite size effects~\cite{Read:2009yp,Hinderer:2009ca,Markakis:2010mp}. Extending our analysis beyond this frequency would only serve to strengthen our results for the following reasons: (i) for a given GW source at a fixed distance, the inclusion of the late inspiral, plunge and merger, increases the SNR, which leads naturally to an improvement in parameter estimation; (ii) the finite size effects that NSs experience can provide useful information in mass extraction and distinguishing between NSs and BHs; and (iii) electromagnetic counterparts from the merger phase can aid in differentiating between NSs and BHs. Thus, from this standpoint, our parameter estimation results could be thought of as conservative. 

\subsection{Model selection}
\label{modelselection}

Given a GW detection, a particularly important follow-up question is whether the signal was produced by a spinning binary or not. In this section, we address this issue by examining whether the restricted, double-precessing model can be used to distinguish between spinning and nonspinning signals. We do so by calculating the~\emph{Bayes factor}, in the case of uninformative flat priors the betting odds, in favor of the spinning model. If the BF is less than one, then the nonspinning model is preferred and we cannot conclude from the data that the binary components have nonzero spin magnitudes. 

When considering nested models, i.e.~models that reduce to each other when a subset of the parameters in one of them acquire certain values, the BF reduces to the {\emph{Savage-Dickey density ratio}}, which is given by
\be
{\rm BF}=\frac{p(\chi_1=0,\chi_2=0)}{p(\chi_1=0,\chi_2=0 | d)}\,,
\ee
the ratio of the prior belief that the spins were zero to the posterior belief that the spins are zero. In Appendix~\ref{app-sd}, we derive this result for models that differ by multiple parameters, some of which do not contribute to the likelihood unless others are nonzero, e.g.~the spin angle parameters do not matter if the spin magnitude is zero.

Although the prior can be easily evaluated at $(\chi_1,\chi_{2})=(0,0)$ since it is uniform, the posterior is much more difficult to calculate. As already mentioned, the process of determining the $2D$ posterior $p(\chi_1,\chi_2 | d)$ involves marginalizing over all other parameters. This is done by dividing the $(\chi_1,\chi_2)$ space into bins of size $d\chi_1=d\chi_2$ and counting how many times the chains visit each corresponding bin. The value of the posterior at $(\chi_1=0,\chi_2=0)$ is proportional to the number of samples in the first bin $(0,d\chi_1,0,d\chi_2)$. 

Clearly, the result depends sensitively on the number of bins used, or equivalently, on the size of each bin. There are two main sources of error in this calculation. If the size of the bins is too small, there will not be enough samples in each of them to give a statistically reliable result, i.e. there are large root $n$ errors, where $n$ is the number of samples in the bin. A very large bin size, on the other hand, will result in an inaccurate estimate of the value of the posterior at $(\chi_1,\chi_{2})=(0,0)$.

\begin{figure}[t]
\begin{center}
\includegraphics[width=\columnwidth,clip=true]{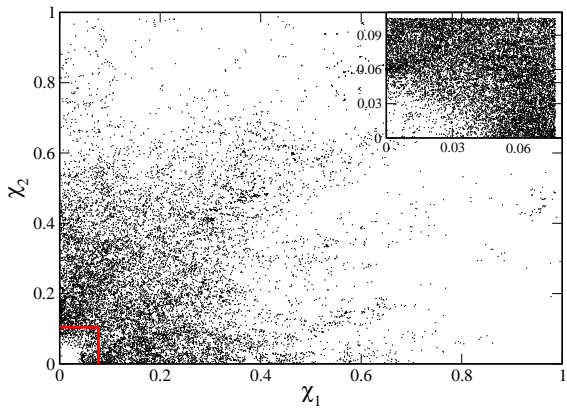}
\caption{\label{fig:chi1-chi2}  (Color online) Example $2$D scatter plots for $\chi_1-\chi_2$ for the pilot run (main plot) and the focused run (inset) to illustrate the two-stage analysis. The red box in the pilot run indicates the size of the region $(0,\chi_1^{\max},0,\chi_2^{\max})$ of the focused run. This region contains $\sim 10 \%$ of the total points of the pilot run. }
\end{center}
\end{figure}

In order to reduce root $n$ error, we need more samples in the first bin, which we achieve through a two-stage analysis. In the first stage, the \emph{pilot run}, we obtain $N_1$ samples from the full posterior distribution. Given that, one can estimate the rectangle $(0,\chi_1^{\max},0,\chi_2^{\max})$ that contains $n_1$ samples ($n_1$ chosen to be $\sim 10\%$ of $N_1$). In the second stage, the \emph{focused run}, we carry out an analysis with a flat prior in $(0,\chi_1^{\max},0,\chi_2^{\max})$ and a zero prior elsewhere, effectively forcing the chains to visit points close to zero spin magnitude, and thus reducing the statistical root $n$ fluctuations. The focused run results in a total of $N_2$ points, $n_2$ of which are in $(0,d\chi_1,0,d\chi_2)$, where recall that $d\chi_1=d\chi_2$ is the size of the bins. Figure~\ref{fig:chi1-chi2} gives an illustration of this procedure. 
The value of the normalized $2D$ posterior at vanishing spins is then
\be
p(\chi_1=0,\chi_2=0 | d) = \frac{n_2}{N_2}\frac{n_1}{N_1}\frac{1}{d\chi_1 d\chi_2}\,,\label{2stageposterior}
\ee
while the fractional error from this procedure can be estimated through
\be
\sqrt{\frac{1}{n_1}+\frac{1}{n_2}}\,.
\label{error-def}
\ee

Having ensured that there are enough samples in the first bins, we still need to choose a bin size that provides an accurate estimate of the value of the posterior at zero spin magnitude. We do so by plotting the posterior in the first bin $(0,d\chi_1,0,d\chi_2)$ as a function of the bin size $d\chi_1=d\chi_2$. From this plot, we choose the points $\{b_i\}$ that satisfy the two following requirements: (i) the bin size is not comparable to the injected spin value and (ii) there are at least $30$-$50$ samples in the first bin. Each point has an error bar $\{b^{\min}_i,b^{\max}_i\}$ calculated through Eq.~\eqref{error-def}. The BF is, then, given by the average of these points with an error bar $\{\min(b^{\min}_i),\max(b^{\max}_i)\}$. The convergence of this procedure, and the accuracy of the error estimates, were checked by performing multiple runs with different random number seeds for a few examples. These multiple runs produced consistent results, with a spread in values that agreed with
the error estimates.

In~\cite{Chatziioannou:2014coa}, we presented the BF in favor of the spinning model as a function of the injected $\chi_{\eff}$ parameter:
\be
\chi_{\eff}=\frac{\vec{\chi}_1 \cdot \hat{L} +\vec{\chi}_2 \cdot \hat{L}}{2},\label{effspin-def}
\ee
for the spin-aligned and the double-precessing models, assuming an aLIGO-AdV injection with SNRs $10$ and $20$. The conclusion from that analysis was that the double-precessing model could state that the signal corresponded to a spinning binary at smaller effective spins than when using the spin-aligned model ($\chi_{\eff}=0.025$ at SNR $=10$ for the former, and $\chi_{\eff}=0.05$ at the same SNR for the latter). 

\begin{figure}[t]
\begin{center}
\includegraphics[width=\columnwidth,clip=true]{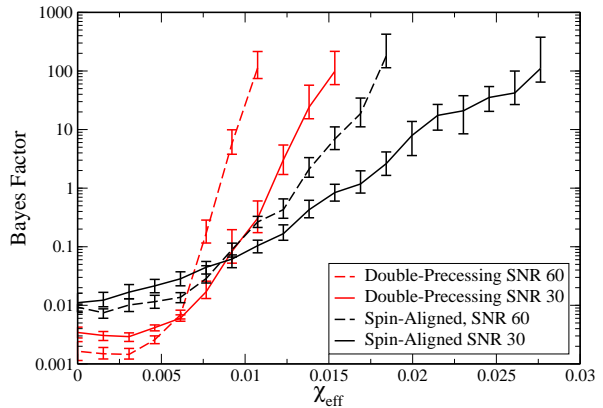}
\caption{\label{fig:BF_NSNS_LIGO3} (Color Online) BF as a function of the injected $\chi_{\eff}$ between nonspinning and spinning models for System $1$ of Table~\ref{systems} and for spin-aligned (black) and double-precessing (red) templates, assuming an injection with SNR $30$ (solid) and $60$ (dashed) as seen by LIGO3. }
\end{center}
\end{figure}

Here we carry out a similar study for a LIGO3 detection. We inject system $1$ of Table~\ref{systems}, where we vary $\chi_1=\chi_2$ and recover it with the double-precessing and the spin-aligned model. Figure~\ref{fig:BF_NSNS_LIGO3} gives the BF in favor of the spinning model as a function of the injected effective spin for a LIGO3 injection with SNR $30$ and $60$. For typical LIGO3 SNRs, the double-precessing model can detect spins as low as $\chi_{\eff}= 0.01$, while the spin-aligned model can do so only above $\chi_{\eff} = 0.02$.
\begin{figure}[t]
\begin{center}
\includegraphics[width=\columnwidth,clip=true]{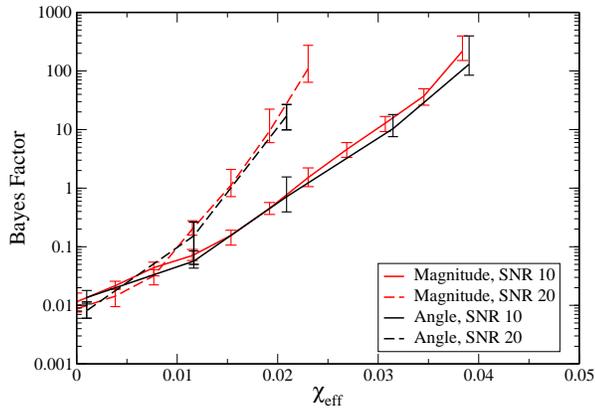}
\caption{\label{fig:BF_comparison} (Color Online) BF as a function of the injected $\chi_{\eff}$ between nonspinning and spinning models with double-precessing templates, where $\chi_{\eff}$ has been updated through a change in the spin magnitudes (red lines, System $1$ of Table~\ref{systems}) and a change in the spin angles (blue lines, System $2$ of Table~\ref{systems}). The injected signal has a SNR of $10$ (solid lines) and $20$ (dashed lines) and is measured by aLIGO. At the same SNR value, both techniques of increasing $\chi_{\eff}$ give similar results, confirming the hypothesis that it is this combination that affects the gravitational waveform.}
\end{center}
\end{figure}
The effective spin parameter is the appropriate variable to use when studying spin detectability~\cite{Chatziioannou:2014coa}. This is because it is $\chi_{\eff}$ which enters to leading PN order in the evolution of the GW phase. We can demonstrate the validity of this argument by calculating the BF in favor of the spinning model as $\chi_{\eff}$ is increased in two different ways: (i) by increasing the value of the injected dimensionless spin parameters $\chi_1 = \chi_{2}$ (System $1$ in Table~\ref{systems}), and (ii) by decreasing the angle between the spin and the orbital angular momenta (System $2$ in Table~\ref{systems}). Figure~\ref{fig:BF_comparison} shows the BF in favor of the spinning model for the double-precessing model calculated in both ways. Both approaches give similar results, demonstrating that the model depends indeed on $\chi_{\eff}$ and not on the individual spin magnitudes and orientations.

\subsection{Accuracy of recovered parameters}
\label{accuracy}

The double-precessing model can break degeneracies between the spin magnitudes and the masses, improving the accuracy of mass extraction significantly, as compared to the spin-aligned model~\cite{Chatziioannou:2014coa}. This is due to the ability of the double-precessing model to better match the complicated likelihood surface of a precessing system thanks to its additional degrees of freedom: the four spin angles. This has nothing to do with the parameter priors associated with each model, uniform in $\chi_1$ and $\chi_2$ for the spin-aligned and uniform in $\chi_1,\chi_2,\cos{\theta_1},\cos{\theta_2},\phi_1$ and $\phi_2$ for the double-precessing model, as we demonstrate in this subsection.

\begin{figure*}[t]
\begin{center}
\includegraphics[width=\columnwidth,clip=true]{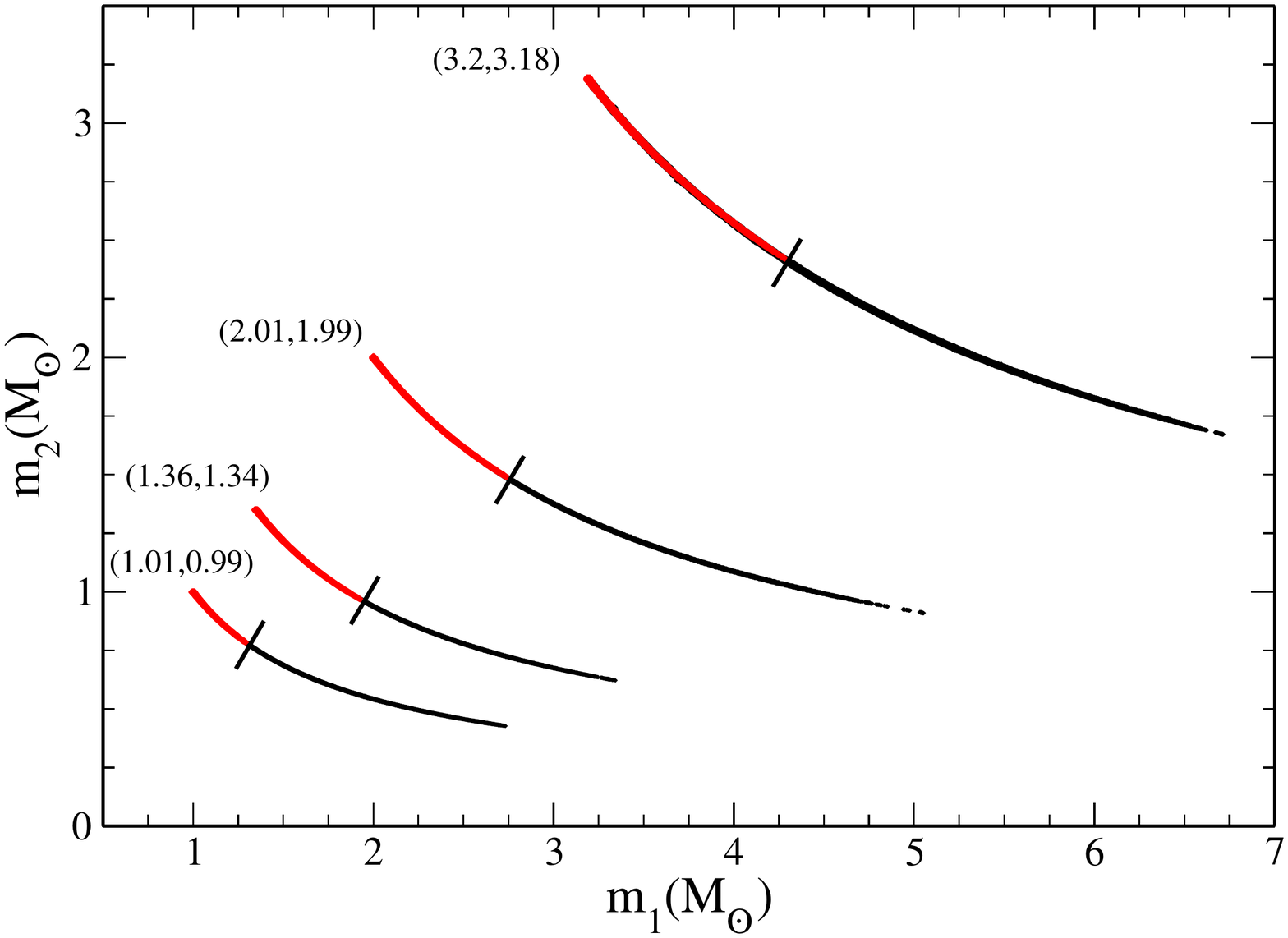}\quad
\includegraphics[width=\columnwidth,clip=true]{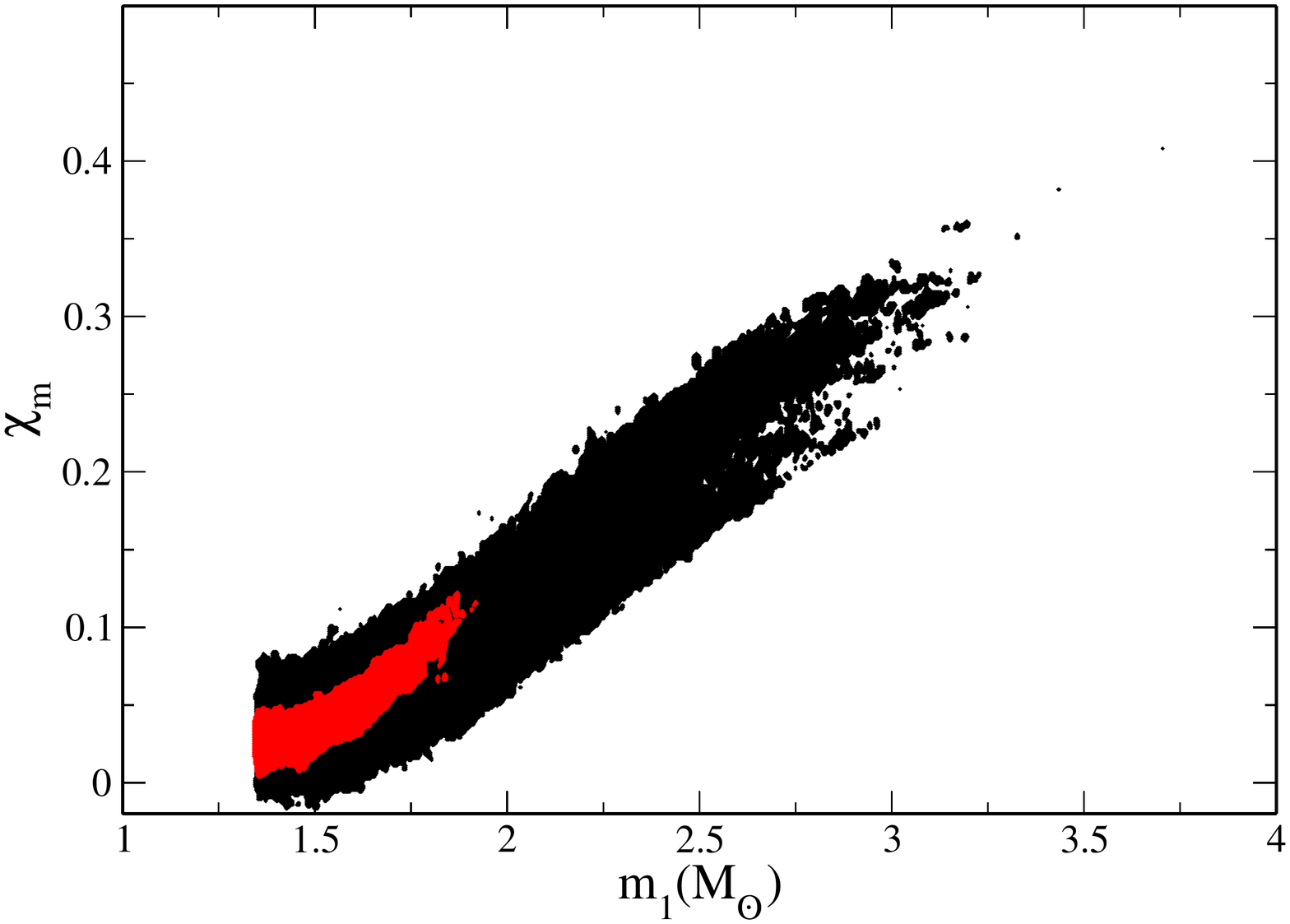}
\caption{\label{fig:scatter_spin004} (Color Online)  (Left panel) Scatter plot of the $90\%$ probability quantile in $(m_1,m_2)$ for System $3$ of Table~\ref{systems} and SNR 10 with different masses extracted with spin-aligned (black) and double-precessing (red) templates. The short lines that cut across the scatter plots mark the boundaries of the quantiles in the direction orthogonal to the chirp mass. (Right panel) Scatter plot showing the $90\%$ probability quantile in the $(\chi_{m}, m_1)$ plane for the $(1.36,1.34) M_\odot$ system
from the left panel. The mass-spin correlation is far more pronounced for the spin-aligned model (black) than for the double-precessing (red) waveform model.}
\end{center}
\end{figure*}

Figure~\ref{fig:scatter_spin004} shows this process at work. The left and right panels show the $90\%$ probability quantile on the $m_1$-$m_2$ and the $\chi_{m}$-$m_{1}$ plane respectively, where $\chi_{m} \equiv (m_1\hat{\chi}_1 \cdot \hat {L} + m_2\hat{\chi}_2 \cdot \hat {L})/(m_1+m_2)$. We use System $3$ of Table~\ref{systems} at SNR $10$, recovered with the double-precessing (red) and the spin-aligned (black) models. The double-precessing model breaks the mass-spin magnitude degeneracy, leading to a much more accurate individual mass determination. A similar figure was shown in~\cite{Chatziioannou:2014coa}, but for nonspinning injections. The results obtained for spinning injections are stronger than for nonspinning ones. 

\begin{figure}[h]
\begin{center}
\includegraphics[width=\columnwidth,clip=true]{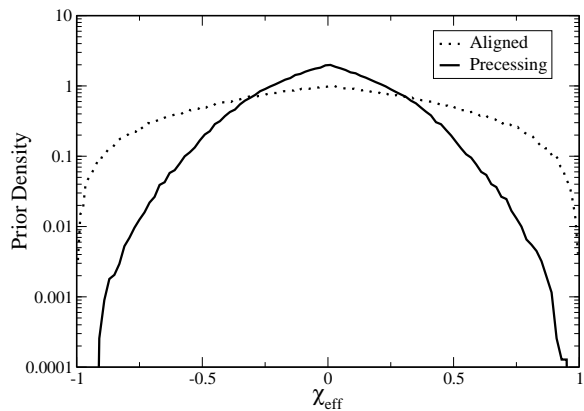}
\caption{\label{fig:chieff_prior} Prior distribution for $\chi_{\eff}$ for the priors we have used in the two different models: uniform spin magnitudes and uniform priors on the unit sphere for all direction angles for the precessing model (solid line), and uniform spin magnitudes for the aligned model (dotted line). In the range of interest $[-0.5,0.5]$ the priors differ by less than a factor of $\sim 3-4$, demonstrating that it is not a difference in priors that results in the increased measurement accuracy of the double-precessing model.}
\end{center}
\end{figure}
Figure~\ref{fig:scatter_spin004}, together with other results in this paper and the results of~\cite{Chatziioannou:2014coa}, demonstrate that the use of the double-precessing model results in a significantly improved parameter extraction accuracy~\emph{even for nonspinning injections}. This might seem counterintuitive, because one may expect that increasing the dimensionality of the parameter space \emph{without} increasing the complexity of the data (as is the case with the nonspinning injection) deteriorates measurement accuracy. Indeed, we find this to be the case when using the spin-aligned model to extract a nonspinning signal; the two extra spin parameters, the spin magnitudes, introduce degeneracies with the individual masses that degrade the accuracy of mass extraction. 

Following that reasoning, one may expect that the double-precessing model would perform even worse, since it has four more parameters than the spin-aligned model: the spin angles. This is \emph{not} the case for the following reason. The spin angles offer the model more ways to leave the region of parameter space where the mass-spin degeneracies are more pronounced. When this occurs, the likelihood calculated between the signal and the double-precessing model deteriorates severely, leading to the rejection of the proposed jumps that have large masses. 

The tendency of the double-precessing model to leave the mass-spin degenerate region of parameter space is~\emph{not} a result of the choice of prior. In the spin-aligned case, we chose uniform priors on the spin magnitudes in the range $[0,1]$. In the double-precessing model, we chose uniform priors on the spin magnitudes in the range $[0,1]$ and uniform priors on the unit sphere for all direction angles. To determine the influence of these choices on our results we imposed the precessing $\chi_{\eff}$ prior on the spin-aligned model. We, indeed, found that the results of Fig.~\ref{fig:scatter_spin004} are not noticeably modified, demonstrating that it is not the choice of prior that enhances the performance of the double-precessing model. This result should not be surprising if one compares the two priors. Figure~\ref{fig:chieff_prior} shows the prior distribution for $\chi_{\eff}$ for the two models considered here.  The two priors on $\chi_{\eff}$ show the same qualitative behavior in the region $[-0.5,0.5]$, the range of interest here (see also the right panel of Fig.~\ref{fig:scatter_spin004}). 

These results indicate that the increased accuracy we achieve with the double-precessing model is~\emph{not} a consequence of our choice of parameter priors, but rather it is due to the likelihood itself and its dependence on the precession features of the signal. The latter offer more ways for the double-precessing model to produce large mismatches with the injected signal, when it has to select the additional four spin angle parameters. As a consequence, the large-spin/large-mass points tend to give lower likelihoods. The double-precessing model prefers to stay in the region of parameter space that fits the injected parameter, rather than wander off into these regions of lower likelihood. Effectively, in nonprecessing models, a change in mass ratio results in a change of the rate of monotonic increase in the phase and the amplitude of the GW. The same can be achieved through a change in spin magnitude, due to the mass-spin degeneracy. On the other hand, precessional effects introduce phase and amplitude modulations that cannot be reproduced by a change in mass ratio. 

This is not the first time that parameter extraction is seen to be greatly improved when a more detailed model is used. Initial studies of the projected bounds of the graviton mass and the Brans-Dicke coupling parameter~\cite{Will:1994fb,Scharre:2001hn,Will:1997bb,Will:2004xi} assumed nonspinning signals. The introduction of aligned spins in the models brought along degeneracies that degraded the bounds by about an order of magnitude~\cite{Berti:2004bd}. However, Stavridis and Will~\cite{Stavridis:2009mb} and Yagi and Tanaka~\cite{Yagi:2009zm} showed that the inclusion of precessional effects in the GW model can bring the bounds back to almost their initial nonspinning values. See~\cite{Yunes:2013dva} for a more detailed discussion of the evolution of these bounds. 

\begin{figure*}[t]
\begin{center}
\includegraphics[width=\columnwidth,clip=true]{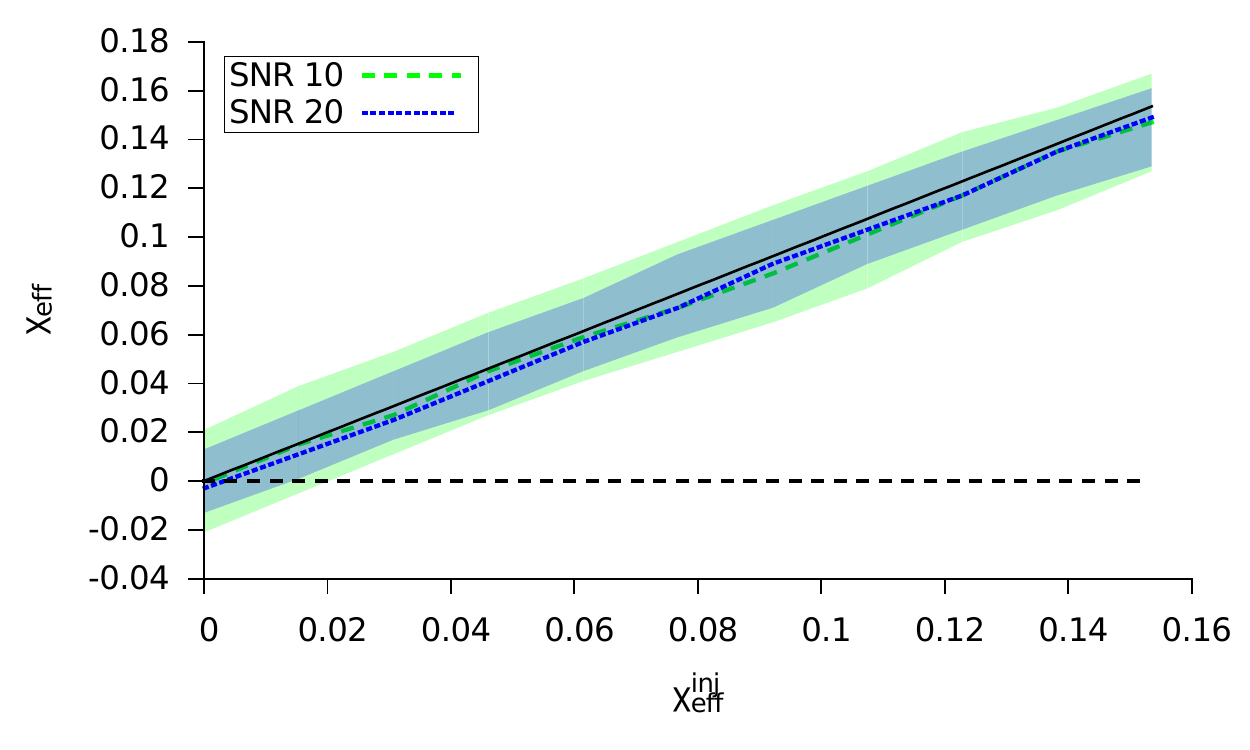} \quad
\includegraphics[width=\columnwidth,clip=true]{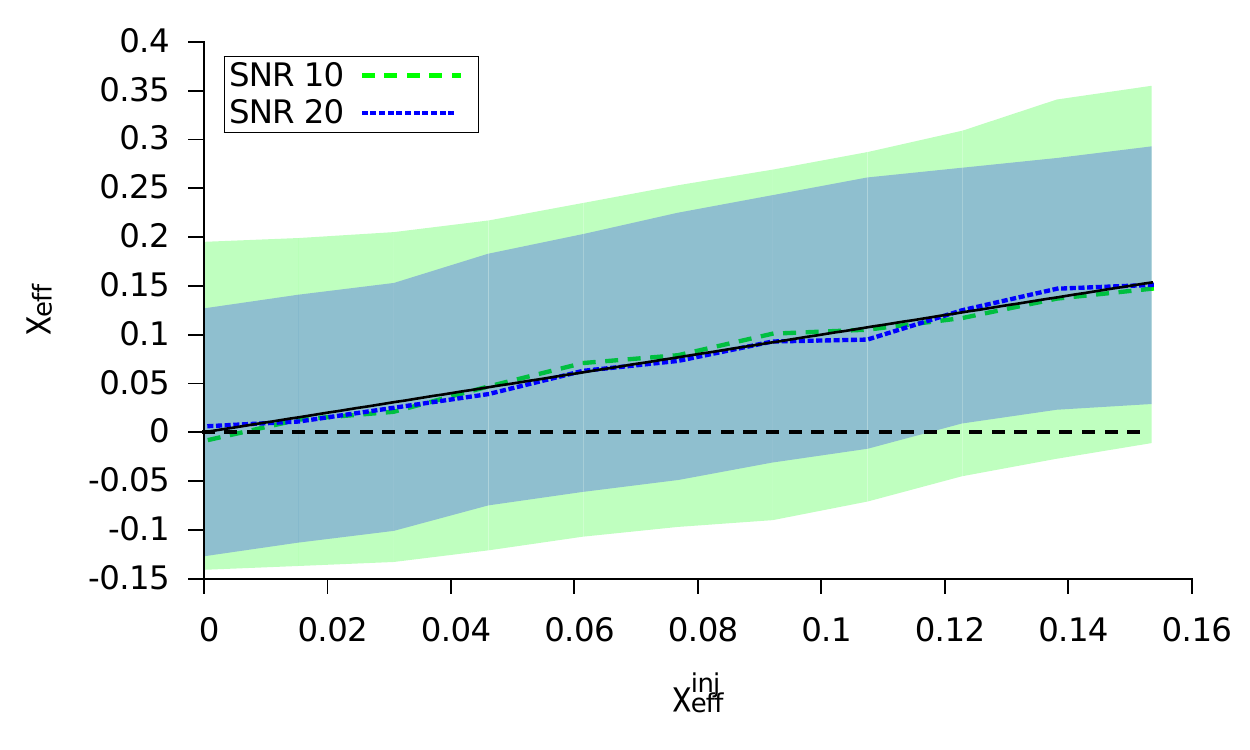} \\
\includegraphics[width=\columnwidth,clip=true]{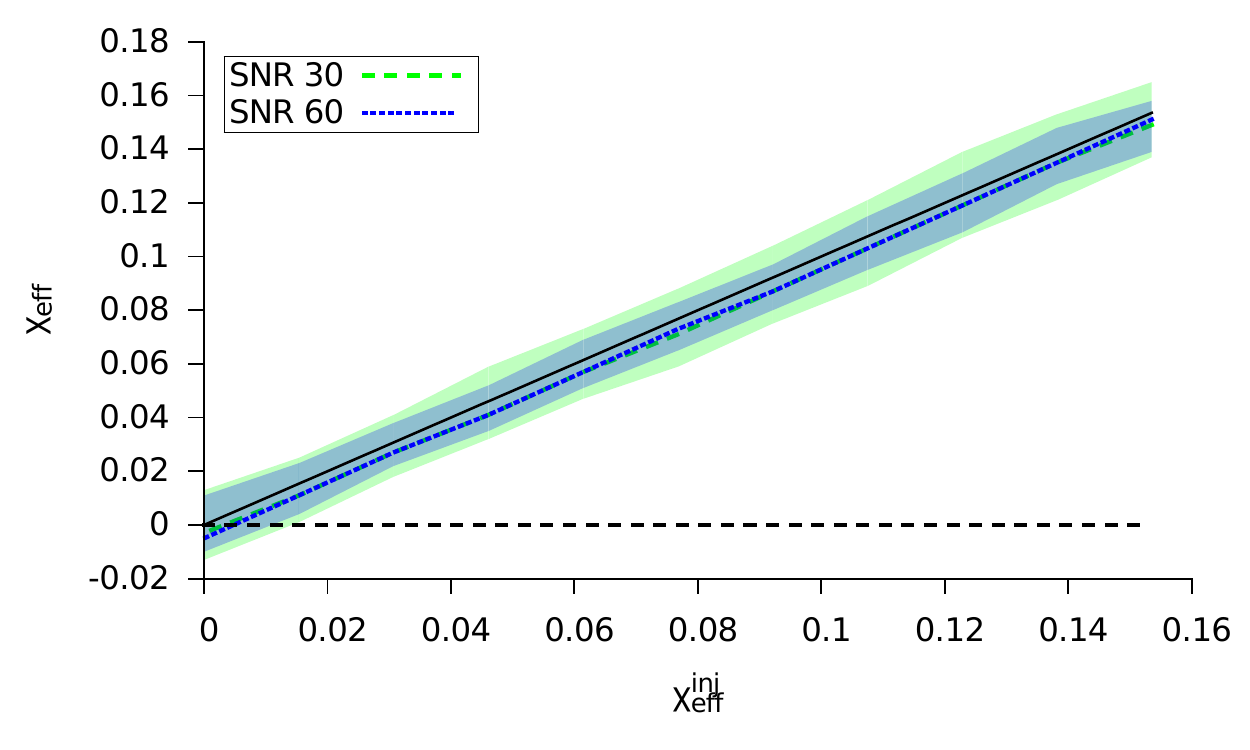} \quad
\includegraphics[width=\columnwidth,clip=true]{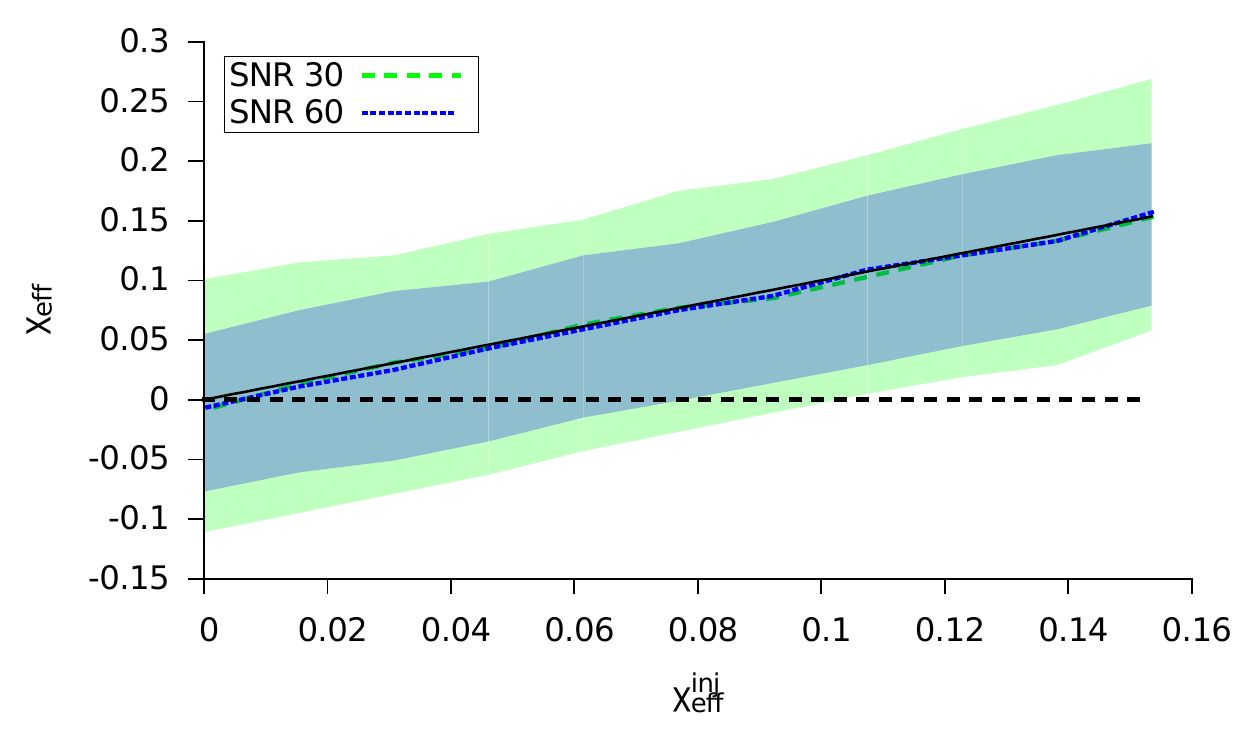} 
\caption{\label{fig:confidence_NSNS} (Color Online) Maximum posterior value for $\chi_{\eff}$ for System $1$ of Table~\ref{systems} as a function of the injected value of $\chi_{\eff}$. The top panels correspond to a signal measured by aLIGO, while the bottom ones are for LIGO3. The left panels show signals recovered with the double-precessing templates, while the right ones with the spin-aligned ones. The shaded regions indicate the minimum interval that contains $68\%$ of the posterior distribution. The black dashed line indicates $\chi_{\eff}=0$, while the black solid line gives the $\chi_{\eff}=\chi_{\eff}^{\inj}$ curve. The use of double-precessing templates reduces the spin extraction error by about an order of magnitude compared to what one would get from spin-aligned templates. Also, for the same model the accuracy is essentially independent  on the injected spin value, and depends mainly on the SNR value.  }
\end{center}
\end{figure*}

Once the components' masses have been accurately recovered, one may also be interested in recovering the spin magnitudes themselves. We quantify our belief on the recovered, effective spin parameter by estimating the minimum interval in the $\chi_{\eff}$ space that contains $68\%$ of the posterior distribution. This interval corresponds to the 1-$\sigma$ confidence region for a Gaussian distribution. Figure~\ref{fig:confidence_NSNS} shows the value of $\chi_{\eff}$ at the peak of the posterior for System $1$ of Table~\ref{systems} as a function of the injected effective spin value for signals detected with aLIGO (top panels) and LIGO3 (bottom panels) with the double-precessing model (left panels) and spin-aligned model (right panels) and SNR of $10$ and $20$ for aLIGO, and $30$ and $60$ for LIGO3. We indicate the $68\%$ confidence area with shaded regions. 

As expected, the distributions are peaked closer to the injected value and the error bars decrease as the SNR increases. The error bars always include the injected value. For an aLIGO detection with a SNR of $10$, one could determine $\chi_{\eff}$ with a confidence of about $\pm 0.02$. The use of spin-aligned templates deteriorates this by an order of magnitude; the error in $\chi_{\eff}$ is now about $\pm 0.2$. We find similar results for signals detected by LIGO3. This time, the double-precessing model can place error bars of about $\pm 0.01$ at SNR $30$, while the spin-aligned model achieves an accuracy of about $\pm 0.1$ only at the same SNR.
%

\section{Conclusions}
\label{conclusions}

We presented a full Bayesian study of the performance of various analytical templates for detection and parameter estimation of double-precessing, compact binary inspirals. We considered the usual nonspinning and spin-aligned models, as well as the new analytical (small-spin), double-precessing model of~\cite{Chatziioannou:2013dza}.

We found that even though the nonspinning and spin-aligned models can be used for the detection of NSNS binaries (symmetric spin parameters up to $0.2$), they are inadequate for BHBH binaries (arbitrary symmetric spin parameters). Furthermore, they induce mismodeling biases that make any use of them for parameter estimation purposes prohibitive. On the other hand, the spin-precessing model can achieve fitting factors above the $98\%$ threshold for all symmetric spin values and lead to a reduction of the systematic bias in mass and spin of at least an order of magnitude~\cite{Vecchio:2003tn,Lang:1900bz,PhysRevD.80.064027,Vitale:2014mka}. This enables us to use it for parameter estimation of NSNS binary systems.

The parameter extraction analysis of this paper is carried through a search of the likelihood surface with a MCMC technique, which is better suited for sampling complicated and weak signals than Fisher information matrix estimates~\cite{Balasubramanian:1995bm,Nicholson:1997qh,Balasubramanian:1997qz,Vallisneri:2007ev,vanderSluys:2008qx,Vitale:2010mr,Vallisneri:2011ts,Rodriguez:2013mla}. We find that the double-precessing template can not only lead to the detection of spins, but also to the measurement of the effective spin parameter to high accuracy compared to an alternative spin-aligned model. The inclusion of precessional effects in the waveforms adds enough information that mass extraction is improved sufficiently to break the degeneracy between NSs and BHs. Reference~\cite{Chatziioannou:2014coa} showed this to be true for nonspinning systems, and here we demonstrate the validity of this conclusion for double-precessing signals. In fact, the addition of spin to the injected signal only serves to strengthen the conclusions of~\cite{Chatziioannou:2014coa}.

The results presented here demonstrate the importance of precessional effects in the analysis of double-precessing systems, even when the spin magnitudes and angles are small. Failure to accurately include them will lead to a significant loss in the volume accessible to GW detectors. Given the already low detection rates expected, such a reduction might lead to erroneous astrophysical conclusions. Apart from the detectability issue, the use of double-precessing templates in data analysis can lead to answers to many astrophysically important questions, like mass and spin distributions of astrophysical objects.

A question that might arise here is whether an unmodeled burst-type search can outperform a template-based search with poor models. The efficiency of burst searches depends on the total mass and the SNR of the signal. For SNRs below 20 and for total masses in the $[20,100]M_{\odot}$ range we expect a poor template to serve better than a burst search when it comes to detection. A full study and quantification of this are left for future work.

A possible candidate source that has been excluded from this analysis is compact binaries of BHs and NSs.  The reason for doing so is that the assumptions of~\cite{Chatziioannou:2013dza}, i.e. small but \emph{comparable} spin magnitudes, break for these systems. The dynamics of BHNS binaries is dominated by the orbital angular momentum and the spin momenta of the BH, with the corresponding orbits and waveforms obeying simple precession~\cite{Apostolatos:1994mx}. One can improve these waveforms perturbatively by including first order corrections in the spin of the NS. 

Another possible way of improving the performance of the waveforms studied here is by performing the Fourier transform of the time-domain waveform in a more robust way. As shown in~\cite{Klein:2013qda}, when precessional effects are large, the assumptions behind the stationary phase approximation are no longer valid. In that case, a more accurate, and complicated, method for the analytical Fourier transform is required. The stationary phase approximation remains valid for spin magnitudes up to $\sim 0.2$. However, here we consider spin values up to $0.98$ and we expect the stationary phase approximation to break for some of these systems~\cite{Klein:2013qda}. Therefore, it is possible that the results showed here could be improved if one carried the Fourier transform through a uniform asymptotics method, similar to that proposed in~\cite{Klein:2013qda}.
\acknowledgments
We would like to thank Laura Sampson for many helpful discussions. We thank Thomas Dent, Mark Hannam, Richard O'Shaughnessy, and Evan Ochsner for comments and suggestions. K.C. acknowledges support from the Onassis Foundation. N.Y. acknowledges support from NSF Grant No. PHY-1114374, NSF CAREER Grant No. PHY-1250636 and NASA Grant No. NNX11AI49G. A. K. and N. C. acknowledge support from the NSF Award PHY-1306702 and NASA Grant No. NNX10AH15G.  

\appendix

\section{Systematic and Statistical Error}
\label{app-errors}

In practice no waveform model will be a perfect match to the signals produced by nature. The mismatch between the true signal and the theoretical model leads to a~\emph{systematic} error in parameter recovery. Furthermore, parameter extraction also suffers from~\emph{statistical} errors arising from noise in the detectors. The systematic errors correspond to a deterministic displacement
away from the true signal manifold, while the statistical errors correspond to a random displacement away from the signal manifold. To leading order, for nearby waveform families, the two types of
error are independent. Let us assume we have two waveform models $h_1$ and $h_2$ that are qualitatively the same, so that we can write
\be
h_2=h_1+\delta h, \label{smalldiff}
\ee
where $\delta h$ is small in the sense $(\delta h \vert h) \ll (h \vert h)$ and $(\delta h \vert \delta h) \ll (h \vert h)$.

The statistical error on the extracted parameters for either model can be approximated through the Fisher information matrix
\be
\Gamma_{ij} = (h_{, i}\vert h_{, j}) \, .
\ee
To leading order in the waveform mismatch we have
\be
\Gamma_{2ij} = \Gamma_{1ij} + (h_{1, i}\vert \delta h_{, j}) + (\delta h_{, i}\vert  h_{1, j}) \, .
\ee
The correction terms will be small if the derivatives of $\delta h$ are also of order $\delta h$. To see this is indeed the case, one can let $h_1$ be the most accurate PN model we can construct (the numerical PN one) and $h_2$ be the double-precessing model of~\cite{Chatziioannou:2013dza}. The difference between the two models is that the latter misses corrections of ${\cal{O}}(\chi^2)$, and thus, $\delta h$ is of $1.5$PN order. To see this, write the waveforms as
\be
h_1= A e^{i\Phi_1},\quad
h_2= A e^{i\Phi_2},
\ee
assuming the amplitudes are the same, since they do not have a large impact from a data analysis point of view. Then, 
\be
\delta h = A (e^{i\Phi_1}-e^{i\Phi_2}) = A (e^{i\Phi_1}-e^{i\Phi_1 + i \delta \Phi}) \sim -A e^{i\Phi_1}i \delta \Phi,
\ee
and its derivative is
\be
\delta h_{,i} \sim -A e^{i\Phi_1}i \delta \Phi_{,i} + A e^{i\Phi_1} \Phi_{1,i} \delta \Phi .
\ee
Since all the derivatives are with respect to the parameters, they do not change the PN order of the terms. Therefore, $\delta h_{,i}\sim \delta h$ and $\Gamma_{2ij} \simeq \Gamma_{1ij}$.

We also present a numerical demonstration of the above result by computing the entries of the Fisher matrix for the two waveform models (the numerical PN and the analytical double-precessing one), evaluated at the same parameters. We do so for the eight physical parameters of interest: the chirp mass ${\cal{M}}$, the total mass $m$, and the six spin parameters $(\theta_1,\phi_1,\chi_1,\theta_2,\phi_2,\chi_2)$ and a system with $(m_1,m_{2},\chi_{1},\chi_{2})=(7.4M_{\odot},6M_{\odot},0.5,0.5)$.  The mismatch between the numerical PN and analytic waveform,
$1-F$, for this system was 0.043. The ratio of the entries for the two matrices $\Gamma^{an}_{ij}/\Gamma^{num}_{ij}$ is
\be
\left({\begin{array}{cccccccc}
 1.01 & 1.02 & 1.05 & 1.06 & 1.01 & 1.01 & 0.97 & 1.04  \\
 1.02 & 1.03 & 1.06 & 1.08 & 1.02 & 1.02 & 0.99 & 1.05  \\
 1.05 & 1.06 & 1.09 & 1.11 & 1.05 & 1.04 & 1.01 & 1.08     \\
 1.06 & 1.08 & 1.11 & 1.12 & 1.07 & 1.06 & 1.03 & 1.09 \\
 1.01 & 1.02 & 1.05 & 1.07 & 1.01 & 1.01 & 0.98 & 1.04	\\
 1.01 & 1.02 & 1.04 & 1.06 & 1.01 & 1.00 & 0.96 & 1.03\\
 0.97 & 0.99 & 1.01 & 1.03 & 0.98 & 0.98 & 0.94 & 1.01	\\
 1.04 & 1.05 & 1.08 & 1.09 & 1.04 & 1.03 & 1.01 & 1.07	
\end{array} }  \right).
\ee
As expected, the fractional difference in the Fisher matrix entries is of order the mismatch.
The similarity of the corresponding entries of the Fisher matrix demonstrates that the mismatch between the numerical PN and the analytical waveform has little effect on the statistical errors calculated here.

\section{Higher-D Savage Dickey Density Ratio}
\label{app-sd}

Let $M_1$ be a simple model nested in the more complex $M_2$. Suppose that the two nested models have a common set of parameters $\vec{\lambda}$.  $M_2$ has additional
amplitudes $\vec{A}$ and angular parameters $\vec{\theta}$. The likelihood for the more complex model $p(d\vert \vec{A},\vec{\theta},\vec{\lambda},M_2)$ reduces
to the likelihood of the simpler model $p(d\vert \vec{\lambda},M_1)$ when $\vec{A}=\vec{0}$. In other words, the angular parameters only affect the
likelihood when the amplitude parameters are nonzero. The evidence for the simpler model is
\begin{equation}
p(d\vert M_1) = \int p(d\vert \vec{\lambda},M_1) p( \vec{\lambda}\vert M_1) d\vec{\lambda},
\end{equation}
and the evidence for the more complex model is
\begin{equation}
p(d\vert M_2) = \int p(d\vert \vec{A},\vec{\theta},\vec{\lambda},M_2) p( \vec{\lambda}, \vec{A}, \vec{\theta}\vert M_2 ) d\vec{\lambda} d\vec{A} d\vec{\theta}.
\end{equation}
The posterior density for model $M_2$ is then
\begin{equation}
p(\vec{\lambda}, \vec{A}, \vec{\theta} \vert d, M_2) = \frac{ p(d\vert \vec{A},\vec{\theta},\vec{\lambda},M_2) p( \vec{\lambda}, \vec{A}, \vec{\theta}\vert M_2 )}{p(d\vert M_2)} .
\end{equation}
Now consider the situation where $p( \vec{\lambda}, \vec{A}, \vec{\theta}\vert M_2 ) = p( \vec{\lambda}) p(\vec{A}) p(\vec{\theta})$ and $p( \vec{\lambda}\vert M_1) = p( \vec{\lambda})$.
The marginal posterior density for model $M_2$ over $\vec{\lambda}$ is then
\begin{equation}
p(\vec{A}, \vec{\theta} \vert d, M_2) = \int p(\vec{\lambda}, \vec{A}, \vec{\theta} \vert d, M_2) \, d\vec{\lambda} ,
\end{equation}
and
\begin{equation}
p(\vec{A}=\vec{0}, \vec{\theta} \vert d, M_2) = \frac{ p(d\vert M_1)p(\vec{A}=\vec{0}) p(\vec{\theta})}{ p(d\vert M_2) },
\end{equation}
where we have used $p(d\vert \vec{A}=\vec{0},\vec{\theta},\vec{\lambda},M_2) = p(d\vert \vec{\lambda},M_1)$. Going a step further, we can marginalize over the angular parameters
to arrive at $p(\vec{A}=\vec{0} \vert d, M_2)  = \int p(\vec{A}=\vec{0}, \vec{\theta} \vert d, M_2) d\vec{\theta}$ which yields
\begin{equation}
p(\vec{A}=\vec{0} \vert d, M_2) = \frac{ p(d\vert M_1)p(\vec{A}=\vec{0})}{ p(d\vert M_2) }.
\end{equation}
We then see that the BF between models 2 and 1, defined as $B_{21} =p(d\vert M_2)/ p(d\vert M_1)$, is given by the Savage-Dicke density ratio
\begin{equation}
B_{21} = \frac{p(\vec{A}=\vec{0})}{p(\vec{A}=\vec{0} \vert d, M_2)}.
\end{equation}
%

\bibliography{review}
\end{document}